\documentclass[3p,12pt]{elsarticle}

\usepackage{cmap} % для кодировки шрифтов в pdf
\usepackage[english,russian]{babel}
\usepackage{citehack}

\usepackage[T2A]{fontenc}
\usepackage[utf8]{inputenc}
\usepackage[small]{caption2}
\usepackage[caption2]{ccaption}% совместимость с caption2
\usepackage{graphicx} % для вставки картинок
\usepackage{color} % названия цветов
\usepackage[colorlinks,linkcolor=black,urlcolor=blue,citecolor=black]{hyperref}%ссылки
\usepackage{ulem}%зачеркивает слова
\usepackage{subfigure}
\usepackage{float}%плавающие объекты там, где надо
\usepackage{amsthm,amsfonts,amssymb,amsmath}

\usepackage{geometry}
\theoremstyle{plain}
\newtheorem{theorem}{Theorem}
\newtheorem{example}{Example}

\newtheorem*{cor}{Corollary}
\newtheorem*{remark}{Remark}
\numberwithin{equation}{section}
\numberwithin{theorem}{section}
\numberwithin{example}{section}
\makeatletter
\def\ps@pprintTitle{%
   \let\@oddhead\@empty
   \let\@evenhead\@empty
   \let\@oddfoot\@empty
   \let\@evenfoot\@oddfoot
}
\makeatother

\begin{document}
\selectlanguage{english}
\def\qed{{\bf \hfill $\Box$}\endtrivlist}
\def\figurename{Fig.}

\begin{frontmatter}



\title{Geometry and Analytics of the Multifacility Weber Problem}

% \author[label1,label2]{Alexei Yu. Uteshev, Elizaveta A. Semenova}
% \address[label1]{cortext command for theassociated footnote}
% \address[label2]{cortext command for theassociated footnote}

%\author{Alexei Yu. Uteshev}
%\author{Elizaveta A. Semenova}

%\address{}

\author{Alexei Yu. Uteshev\corref{mycorrespondingauthor}}
\ead{alexeiuteshev@gmail.com}
\author{Elizaveta A. Semenova}
\ead{semenova.elissaveta@gmail.com}
\address{Faculty of Applied Mathematics, St. Petersburg State University \\
7/9 Universitetskaya nab., St. Petersburg, 199034 Russia}
\cortext[mycorrespondingauthor]{Corresponding author}
%\ead{alexeiuteshev@gmail.com}


%\author[mysecondaryaddress]{1Alexei Yu. Uteshev\corref{mycorrespondingauthor}}
%\ead{alexeiuteshev@gmail.com}

%\author[mymainaddress,mysecondaryaddress]{2Elizaveta A. Semenova}
%\ead[url]{sem.com}

%\address[mymainaddress]{11}
%\address[mysecondaryaddress]{22}

\begin{abstract}
For the Weber problem of construction of  the minimal cost planar weighted network connecting four terminals with
two extra facilities, the solution by radicals is proposed. The conditions for existence of the network in the assumed topology
and the explicit formulae for coordinates of the facilities are presented. The obtained results are utilized for investigation of the
network dynamics under variation of parameters. Extension of the results to the general Weber problem is also discussed.
\end{abstract}


\end{frontmatter}

%%%%%%%%%%%%%%%%%%%%%%%%%%%%%%%%%%%%%%%%%%%%%%%%%%%%%%
%%%%%%%%%%%%%%%%%%%%%%%%%%%%%%%%%%%%%%%%%%%%%%%%%%%%%%
%  INTRO
%%%%%%%%%%%%%%%%%%%%%%%%%%%%%%%%%%%%%%%%%%%%%%%%%%%%%%
%%%%%%%%%%%%%%%%%%%%%%%%%%%%%%%%%%%%%%%%%%%%%%%%%%%%%%

\section{Introduction}

The classical \textit{Weber} or \textit{generalized Fermat-Torricelli problem} is stated as that of finding the point (facility, junction) $ W=(x_{\ast},y_{\ast}) $  that minimizes the sum of weighted distances  from itself to $ n \ge 3 $ fixed points (terminals) $ \{ P_j=(x_j,y_j)\}_{j=1}^n $ in the Euclidean plane:
\begin{equation}
\min_{W\in \mathbb R^2} \sum_{j=1}^n m_j |WP_j| \, .
\label{Weber_uni}
\end{equation}
Hereinafter $ | \cdot | $ stands for the Euclidean distance and the \textit{weights} $ \{m_j\}_{j=1}^n $ are assumed to be positive real numbers.

The treatment of the problem in the case $ n=3 $ terminals was first undertaken in 1872 by Launhardt \cite{Launhardt} whose interest stemmed from the evident relation to the Economic Geography problem nowadays known as \textit{Optimal Facility Location}. For instance, one can be interested in minimizing transportation costs for a plant
manufacturing one ton of the final product from $ \{m_j\}_{j=1}^n $ tons of distinct raw materials located at corresponding $ \{ P_j \}_{j=1}^n $.

Further development of the problem was carried out in 1909 by Alfred Weber. First, he suggested a different economic interpretation for the three-terminal problem. Let $ P_3 $ be a place of consumption of $ m_3 $ tons of a product produced from two different types of raw materials:  $ m_1 $ tons of the first type located at $ P_1 $ and $ m_2 $ tons of the second type located at $ P_2 $, let $ m_3 < m_1+m_2 $. Where is the optimal location of the production?
In the course of the economic background, Weber formulated the following extension of the problem to the case of $ 4 $ terminals\footnote{In the following citation we change the original notation of the points.}

``Let us take a simple case, an enterprise with three material deposits and one which is capable of being split, technologically speaking, into two stages. In the first stage two materials are combined into a half-finished product; in the second stage this half-finished product is combined with the third material into the final product\dots Let us suppose that possible location of the split production would be in $ W_1 $ and $ W_2 $; $ W_1 $ for the first stage and $ W_2 $ for the second stage. What will be the result if the splitting occurs?''\cite{Weber}

Mathematically the stated problem can be formulated as that of finding the points $ W_1=(x_{\ast},y_{\ast}) $ and $ W_2=(x_{\ast \ast},y_{\ast \ast}) $ which yield
\begin{eqnarray}
& \displaystyle \min_{\{W_1,W_2\} \subset \mathbb R^2} F(W_1,W_2)  \quad \mbox{ where } & \nonumber \\
& F(W_1,W_2)= m_1|W_1P_1|+m_2|W_1P_2| +m_3|W_2P_3|+m_4|W_2P_4|+ m |W_1W_2|
\label{F_Weber}
\end{eqnarray}
and the weights $ \{ m_j\}_{j=1}^4, m $ are treated as given positive real numbers.

The general  \textit{Multifacility Weber problem} is stated as that of location of the  given number  $ \ell \ge 2 $ of the facility points (or, simply, facilities) $ \{W_{i}\}_{i=1}^{\ell} $ in $ \mathbb R^d $ connected to the terminals $ \{ P_j\}_{j=1}^n \subset \mathbb R^d $ that solve the optimization problem
\begin{eqnarray}
\min_{\{W_{1},\dots, W_{\ell}\}\subset \mathbb R^d} \left\{ \sum_{j=1}^n \sum_{i=1}^{\ell} m_{ij} |W_iP_j| +
\sum_{k=1}^{\ell} \sum_{i=k+1}^{\ell-1} \widetilde m_{ik} |W_iW_k|
\right\} \, ;
\label{F_Weber_m}
\end{eqnarray}
here some of the weights $ m_{ij} $ and $ \widetilde m_{ik} $ might be zero.
We will refer to this value  as to the \textbf{minimal cost of the network}.
This problem can be considered as a natural generalization of the celebrated \textit{Steiner minimal tree problem} aimed at construction of the network of minimal length linking the given terminals.
Dozens of papers are devoted to the Weber problem, its ramifications and applications; we refer to \cite{DKSW,ReVEis,Xue1994} for the reviews. The majority of them are concerned with the problem statement where the objective function (\ref{F_Weber_m}) is free of the inter-facilities connections, i.e. all the  weights $  \widetilde m_{ik} $ are zero. This problem is known as the \textit{Multisource Weber problem} or the $ p $\textit{-median problem}\footnote{With $ p $  standing for the number of facilities.}. The present paper is focused on solution to the Multifacility Weber problem.
The mainstream approach in the treatment of this nonlinear optimization problem is the one based on reducing it to an appropriate iterative  numerical procedure. For instance, the unifacility version of the problem (\ref{Weber_uni}) can be resolved via the modified  Weiszfeld algorithm.
The main obstacle of this approach consists in the fact that the objective (or cost) function of the Weber problem is non-differentiable at terminal points, and the iterative procedure might diverge if any of the facilities happens to lie close to a terminal (or, in case of the multifacility problem, if two facilities are about to collide).

The present paper is devoted to an alternative  approach for the  problem, namely an analytical one. We are looking for the conditions for existence of the network and the explicit expressions for the  facility coordinates in terms of the problem parameters, i.e. terminal coordinates and weights. This approach has been originated in the recent papers \cite{Uteshev_GFT} and \cite{Uteshev_Steiner} where the unifacility Weber problem for three terminals and the (full) Steiner minimal tree problem for four terminals had been solved \textit{by radicals}.
Within the framework of this approach, we will focus here on solution to the planar multifacility Weber problem for the case of $ n=4 $  terminals and $ \ell=2 $ facilities (i.e. the problem  (\ref{F_Weber})), and also for the case of $ n=5 $ terminals and $ \ell=3 $ facilities.

Our analytical treatment stems from geometric solution to the problem originated by Georg Pick and published in the Mathematical Appendix   of Weber's book \cite{Weber}. Pick's solution, which we trace in Section \ref{SGeo}, can be interpreted as a counterpart of the algorithm worked out  by Gergonne in 1810  (and rediscovered by Melzak in 1961) for solution of the Steiner minimal tree problem for four terminals.
Nevertheless, Pick did not provide any proof of validity for his algorithm. We also failed to find any references to Pick's solution in subsequent papers on the subject. In the conference paper \cite{Uteshev_Semenova} the present authors have announced without a proof the claim that the Weber problem (\ref{F_Weber}) is solvable by radicals. In a simplified version (and with an extra assumption missed in \cite{Uteshev_Semenova}), this statement is now proved in Section \ref{SAn}.  The deduced formulae for the facilities coordinates approve analytically Pick's considerations.  In addition, the conditions for the existence of the desired configuration of the network are provided.

In the case of the problem involving variable parameters, analytics provides one with a unique opportunity to evaluate their influence on the solution. In particular, it gives the means to determine the  \textit{bifurcation values} for these parameters, i.e. those responsible for the degeneracy of the network topology.
We discuss these issues in Section \ref{SPar} via investigation of the facilities dynamics under
variation of the terminals location or the value of the involved weights\footnote{One may imagine a relevant economic optimization problem with a trawler fishing in the ocean and a
floating fish processing facility drifting in anticipation of the catch transferred to it.}.
 We also prove here that, in the case of existence, the optimal bifacility network has its cost lower than the unifacility one.
Based of the deduced analytical formulae  we suggest in Section  \ref{S=4Alt}, as an alternative to Pick's construction,  geometrical solution which is more attractive than the latter when dealing with the wandering terminal case.

In Section \ref{S>4}, we briefly discuss an opportunity for extension of the results to the case of $ n\ge 5 $ terminals and $ \ell\ge 3 $ facilities. This extension is based on the reduction of the problem to a similar one with $ n-1 $ terminals and  $ \ell-1 $  facilities via a replacement of a pair of terminals by a suitable auxiliary \textit{phantom} terminal. This trick is just a counterpart of the one utilized in Melzak's algorithm for Steiner tree construction.

%%%%%%%%%%%%%%%%%%%%%%%%%%%%%%%%%%%%%%%%%%%%%%%%%%%%%%
%%%%%%%%%%%%%%%%%%%%%%%%%%%%%%%%%%%%%%%%%%%%%%%%%%%%%%
%  3 TERMINALS
%%%%%%%%%%%%%%%%%%%%%%%%%%%%%%%%%%%%%%%%%%%%%%%%%%%%%%
%%%%%%%%%%%%%%%%%%%%%%%%%%%%%%%%%%%%%%%%%%%%%%%%%%%%%%

\section{Unifacility case}\label{S3Term}

\subsection{Three terminals}

We first outline the geometric approach to the problem given in the paper by Launhardt \cite{Launhardt}.

\begin{example}\label{ft0}
Find the optimal position of the facility $ W $ to the problem (\ref{Weber_uni}) where
$$
\left\{\begin{array}{c|c|c}
P_1=(1,5) & P_2=(2,1) & P_3=(7,2)  \\
m_1=3 & m_2=2 & m_3=3
\end{array} \right\}.
$$
\end{example}

\begin{figure}[H]\center
{
\begin{minipage}[t]{130mm}
\begin{minipage}[t]{60mm}
\graphicspath{{Illustrations/}}
\includegraphics[width=45mm]{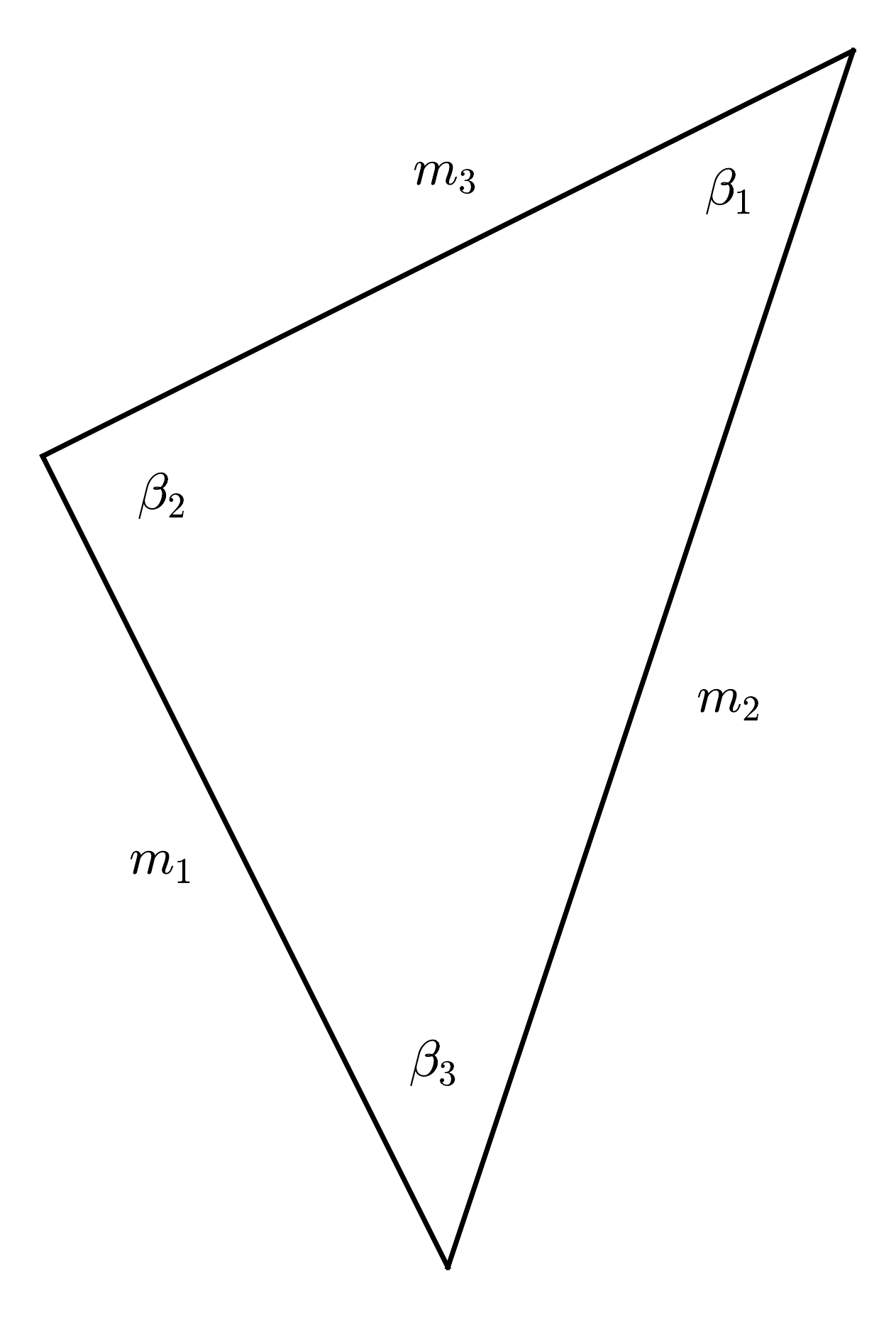}
\begin{center}
		(a)
\end{center}
\end{minipage}
\hfill
\begin{minipage}[t]{60mm}
\graphicspath{{Illustrations/}}
\includegraphics[width=60mm]{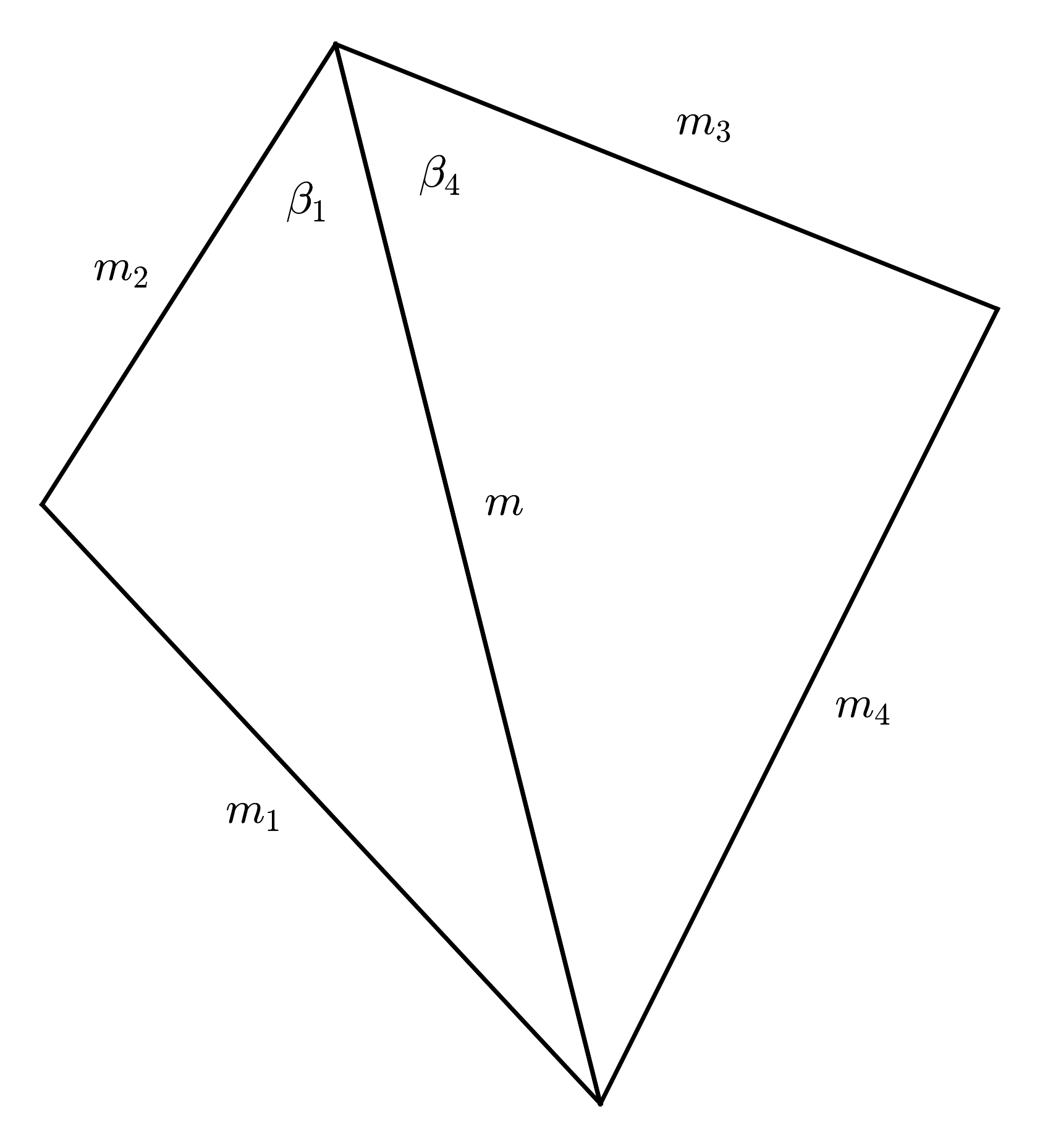}
	
\begin{center}
		(b)
\end{center}
\end{minipage}
\end{minipage}
}
\caption{Weight (a) triangle and (b) quadrilateral}  \label{fig:WeightQuadr}
\end{figure}

\textbf{Solution.} First find the point $ Q_1 $ lying on the opposite side of the line $ P_1P_2 $ with respect to the point $ P_3 $ and such that
$$
 |P_1Q_1|=\frac{m_2}{m_3}|P_1P_2|, \ |P_2Q_1|=\frac{m_1}{m_3}|P_1P_2| \, .
$$
This condition means that the triangle $ P_1P_2Q_1 $ is similar to the so-called \textbf{weight triangle} of the problem, i.e. the triangle composed of the sides formally coinciding with the values of the weights $ m_1, m_2, m_3 $. We will further denote this triangle by $ \{ m_3,m_1,m_2  \} $ (Fig. \ref{fig:WeightQuadr} (a)).

Next, draw the circle $ C_1 $ circumscribing $P_1P_2Q_1$. Finally draw the line through $ Q_1 $ and $ P_3 $. The intersection point of this line with $ C_1 $ is the position of the optimal facility $ W $.
The corresponding (minimal) cost is equal to $ m_3|P_3Q_1| $.  \qed

The feasibility of the suggested algorithm evidently depends on the condition for the existence of the weight triangle, i.e. $ m_3<m_1+m_2, m_1<m_2+m_3,m_2<m_1+m_3 $. However, even under this assumption, the locus of the point $ W $ might lie outside of the triangle $ P_1P_2P_3 $, and, in this case, the obtained solution contradicts the common sense.

This geometrical solution can be developed to an analytical one \cite{Uteshev_GFT}.

\begin{theorem} \label{Th3term1} Denote by $ \alpha_1, \alpha_2, \alpha_3 $ the angles of the triangle $ P_1P_2P_3 $, while by $ \beta_1, \beta_2, \beta_3 $ the angles of the weight triangle (according with a rule similar to that displayed in Fig. \ref{fig:WeightQuadr} (a)). The necessary and sufficient condition for the existence of solution to the problem
\begin{equation}
 \min_{W\in \mathbb R^2} (m_1 |WP_1|+ m_2 |WP_2| + m_3 |WP_3|)
 \label{3termW}
\end{equation}
is that of the following system of inequalities
$$ \{\cos \alpha_j +  \cos \beta_j>0 \}_{j=1}^3 \, . $$
Under this condition, the coordinates of the optimal facility $ W=(x_{\ast},y_{\ast}) $ are given by the formulae
$$
x_{\ast}=\frac{K_1K_2K_3}{2 |\mathfrak S| \sqrt{\mathbf k} d} \left(\frac{x_1}{K_1}+\frac{x_2}{K_2}+\frac{x_3}{K_3} \right), \
y_{\ast}=\frac{K_1K_2K_3}{2 |\mathfrak S| \sqrt{\mathbf k} d} \left(\frac{y_1}{K_1}+\frac{y_2}{K_2}+\frac{y_3}{K_3} \right)
$$
with the cost of the optimal network
$$ \mathfrak C= \sqrt{d} \, . $$
Here
$$
d=\frac{1}{\sqrt{\mathbf k}} (m_1^2K_1+m_2^2K_2+m_3^2K_3)
$$
$$
= |\mathfrak S| \sqrt{\mathbf k} + \frac{1}{2}\left[m_1^2(r_{12}^2+r_{13}^2-r_{23}^2)+
m_2^2(r_{23}^2+r_{12}^2-r_{13}^2)+m_3^2(r_{13}^2+r_{23}^2-r_{12}^2) \right] \ ,
$$
$$
r_{ij}=|P_iP_j|=\sqrt{(x_i-x_j)^2+(y_i-y_j)^2} \quad for \ \{i,j\} \subset \{1,2,3\} \ ,
$$
$$
\mathfrak S=x_1y_2+x_2y_3+x_3y_1-x_1y_3-x_3y_2-x_2y_1 \ ,
$$
\begin{equation}
\mathbf k= (m_1+m_2+m_3)(-m_1+m_2+m_3)(m_1-m_2+m_3)(m_1+m_2-m_3) \ ,
\label{k}
\end{equation}
and
$$
\left\{
\begin{array}{ccc}
K_1&=&(r_{12}^2+r_{13}^2-r_{23}^2)\sqrt{\mathbf k}/2+(m_2^2+m_3^2-m_1^2) |\mathfrak S| , \\
K_2&=&(r_{23}^2+r_{12}^2-r_{13}^2)\sqrt{\mathbf k}/2+(m_1^2+m_3^2-m_2^2) |\mathfrak S| , \\
K_3&=&(r_{13}^2+r_{23}^2-r_{12}^2)\sqrt{\mathbf k}/2+(m_1^2+m_2^2-m_3^2) |\mathfrak S| .
\end{array}
\right.
$$
\end{theorem}

The proof consists in formal verification of the equalities
\begin{eqnarray}
m_1 \, \frac{x_{\ast}-x_1}{|WP_1|}+m_2  \frac{x_{\ast}-x_2}{|WP_2|}+m_3 \, \frac{x_{\ast}-x_3}{|WP_3|}&=&0,
\label{grad_32} \\
m_1 \, \frac{y_{\ast}-y_1}{|WP_1|}+m_2  \,\frac{y_{\ast}-y_2}{|WP_2|}+m_3 \, \frac{y_{\ast}-y_3}{|WP_3|}&=&0, \label{grad_322}
\end{eqnarray}
providing the stationary points of the objective function $ \sum_{j=1}^3 m_j |P_j W| $.

The theorem states that the three-terminal Weber problem is solvable by radicals.
Geometric meaning of the constants appeared in this theorem is as follows: $ \frac{1}{2} |\mathfrak S | $ equals the area of the triangle $ P_1P_2P_3 $ while $ \frac{1}{4}\sqrt{\mathbf k} $ equals (due to Heron's formula) the area of the weight triangle.

We now formulate two technical results to be exploited later. They can be proved via formal application of Theorem \ref{Th3term1}.

\begin{theorem} \label{Th3term2} If the facility $ W $ is the solution to the problem (\ref{3termW}) for some configuration \break $ \renewcommand{\arraystretch}{0.5} \left\{ \begin{array}{c|c|c} P_1 & P_2 & P_3  \\ m_1 & m_2 & m_3  \end{array} \right\} $  then this facility remains unchanged for the  con\-fi\-gu\-ration $ \renewcommand{\arraystretch}{0.5}  \left\{ \begin{array}{c|c|c} P_1 & P_2 & \widetilde P_3  \\ m_1 & m_2 & m_3 \end{array} \right\}  $
with any position of the  terminal $ \widetilde P_3 $ in the half-line $ WP_3 $.
\end{theorem}

\begin{theorem} \label{Th3term3}
For any position of the terminal $ P_3 $, the facility $ W $ lies in the arc of the circle $ C_1 $ passing through the points $ P_1, P_2 $ and
%$$ Q_1=\left(\frac{1}{2}(x_1+x_2)+\frac{(m_1^2-m_2^2)(x_1-x_2)-\sqrt{\mathbf k}(y_1-y_2)}{2m_3^2},
%    \frac{1}{2}(y_1+y_2)+\frac{(m_1^2-m_2^2)(y_1-y_2)+\sqrt{\mathbf k}(x_1-x_2)}{2m_3^2} \right) $$
\begin{eqnarray}
Q_1&=&\Bigg(\frac{1}{2}(x_1+x_2)+\frac{(m_1^2-m_2^2)(x_1-x_2)-\sqrt{\mathbf k}(y_1-y_2)}{2m_3^2}, \nonumber \\
    & &\hspace{20mm}\frac{1}{2}(y_1+y_2)+\frac{(m_1^2-m_2^2)(y_1-y_2)+\sqrt{\mathbf k}(x_1-x_2)}{2m_3^2} \Bigg) \, .
\label{Q1}
\end{eqnarray}
Its center is at
$$  \left( \frac{1}{2}(x_1+x_2)-\frac{m_1^2+m_2^2-m_3^2}{2 \sqrt{\mathbf k}}(y_1-y_2),
         \frac{1}{2}(y_1+y_2)+\frac{m_1^2+m_2^2-m_3^2}{2 \sqrt{\mathbf k}}(x_1-x_2)  \right)
$$
while its radius equals $ m_1m_2 |P_1P_2|/\sqrt{\mathbf k} $.
\end{theorem}

\begin{theorem} \label{Th3term31} Let the terminals $ P_1,P_2,P_3 $ be counted counterclockwise and the
conditions of Theorem \ref{Th3term1} be satisfied. Set
\begin{equation}
\mathfrak S_1=
\left| \begin{array}{rrr}
1 & 1 & 1 \\
x & x_2 & x_3 \\
y & y_2 & y_3
\end{array}
\right|, \
\mathfrak S_2=
\left| \begin{array}{rrr}
1 & 1 & 1 \\
x_1 & x & x_3 \\
y_1 & y & y_3
\end{array}
\right|,\
\mathfrak S_3=
\left| \begin{array}{rrr}
1 & 1 & 1 \\
x_1 & x_2 & x \\
y_1 & y_2 & y
\end{array}
\right| \, .
\label{S1S2S3}
\end{equation}
For any value of the weight $ m_3 $, the optimal facility $ W $ lies in the arc of the algebraic curve of the 4th degree given by the equation
\begin{equation}
m_1^2 \mathfrak S_2^2 \left[(x-x_2)^2+(y-y_2)^2\right] = m_2^2 \mathfrak S_1^2 \left[(x-x_1)^2+(y-y_1)^2\right] \, .
\label{S3m}
\end{equation}
\end{theorem}

\textbf{Proof.} If the conditions of Theorem \ref{Th3term1} are fulfilled then the coordinates of the optimal facility $ W=(x_{\ast},y_{\ast}) $ satisfy the system (\ref{grad_32})--(\ref{grad_322}). Treating this system as linear with respect to $ m_1,m_2,m_3 $, one
arrives at the following relation
$$ m_1 : m_2 : m_3=|WP_1| \mathfrak S_1 : |WP_2| \mathfrak S_2 : |WP_3| \mathfrak S_3 \, .
$$
\qed

\begin{example}\label{ex3t1f0}
For the configuration
$$
\left\{\begin{array}{c|c|c}
P_1=(1,5) & P_2=(2,1) & P_3=(7,2)  \\
m_1=3 & m_2=2 & m_3
\end{array} %\mbox{ and }
\right\} \, ,
$$
find the locus of the facility $ W $ under variation of the weight $ m_3 $.
\end{example}

\begin{figure}[H]\center
{
\graphicspath{{Illustrations/}}
\includegraphics[scale=0.4]{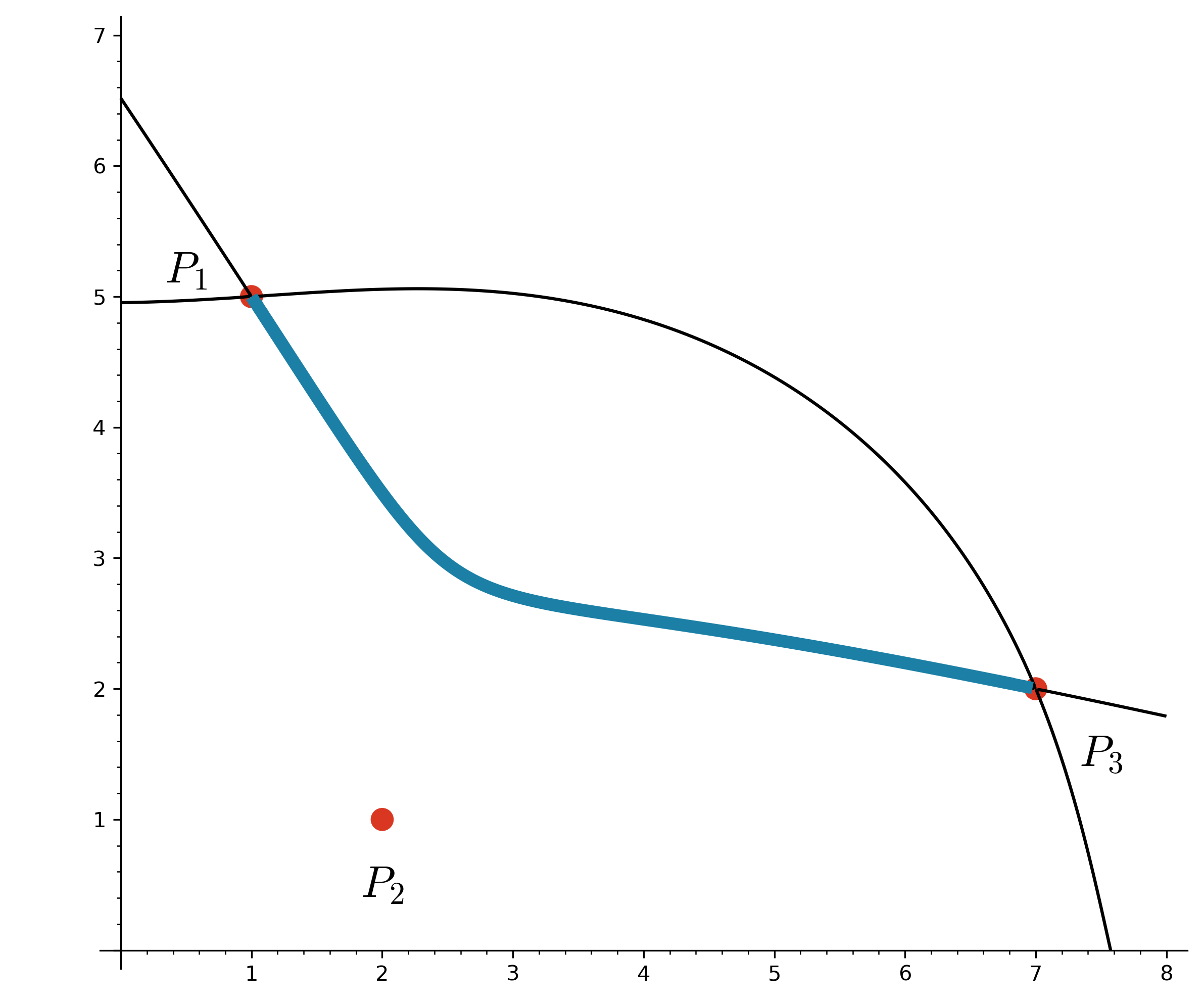}
   % \rule{4cm}{3cm}
%    \label{fig:curve_ed_3_1}
}
\caption{Example \ref{ex3t1f0}. Optimal facility location under variation of the weight $ m_3 $} \label{weber_3_1}
\end{figure}

\textbf{Solution.} Equation (\ref{S3m}) takes the form
$$
7(11\,x+8\,y)(x+4\,y)(x^2+y^2)-(2122\,x^3+4942\,x^2y+3950\,xy^2+3092\,y^3)
$$
$$
+17242\,x^2+20480\, xy+14713\, y^2-48666\, x-33942\, y+48069=0 \, .
$$
The curve is depicted in Fig. \ref{weber_3_1}, with its only branch giving the position of $ W $ displayed in blue boldface.
The values of $ m_3 $ corresponding to this branch lie within the interval
$$ \left[\frac{1}{85}(\sqrt{48365}-12 \sqrt{85}),\ \frac{1}{65}\sqrt{54925+3510\sqrt{130}}   \right] \approx [ 1.285716, 4.740488] $$
with its ends corresponding to collision of the optimal facility with $ P_1 $ or $ P_3 $.  \qed

At the end of the present subsection, we formulate a result that looks somewhat unexpected.

\begin{theorem} \label{Thdual} Construct the point similar to (\ref{Q1}) for every side of the triangle $ P_1P_2P_3 $. Redesignate the point $ Q_1 $ by $ \widetilde Q_3 $, and denote by
$ \widetilde Q_1 $ the point  obtained from $ \widetilde Q_3 $ via cyclic substitution
$$ (x_1,y_1) \to (x_2,y_2), (x_2,y_2) \to (x_3,y_3), m_1\to m_2,m_2\to m_3, m_3 \to m_1 \, . $$
Let $ \widetilde  Q_2 $ be obtained from $ \widetilde Q_1 $ in a similar manner.
For the configurations
$$
\left\{\begin{array}{c|c|c}
P_1 & P_2 & P_3  \\
m_1 & m_2 & m_3
\end{array} %\mbox{ and }
\right\} \quad \mbox{ and } \quad
\left\{\begin{array}{c|c|c}
\widetilde Q_1 & \widetilde Q_2 & \widetilde Q_3  \\
m_1 & m_2 & m_3
\end{array} %\mbox{ and }
\right\} \, ,
$$
locations of the optimal facilities for the problem (\ref{3termW})
are identical. The costs of the corresponding optimal networks are connected by the equality
$$ \widetilde{\mathfrak C} = 2 \mathfrak C \, . $$
\end{theorem}

\subsection{Four terminals}

\textbf{Assumption 1.} Hereinafter we will treat the case where the terminals $ \{ P_j\}_{j=1}^4 $, while counted counterclockwise, compose a convex quadrilateral $P_1P_2P_3P_4$.

Stationary points of the function $ \sum_{j=1}^4 m_j |WP_j| $ are given by the system of equations
\begin{equation}
\sum_{j=1}^4 \frac{m_j(x-x_j)}{|WP_j|}=0,\ \sum_{j=1}^4 \frac{m_j(y-y_j)}{|WP_j|}=0 \, .
\label{StatPoint4}
\end{equation}
Though this system is not an algebraic one with respect to $ x,y $, it can be reduced to this form via successive squaring of every equation. This permits one to apply the procedure of elimination of a variable via computation of the \textbf{resultant}. Thereby, the problem of finding the coordinates of the facility $ W $ can be reduced to that of resolving a univariate algebraic equation \cite{Uteshev_FTC}. Unfortunately, for the considered case, this equation is generically of degree $ 10 $ (v. Example \ref{ex3} below) and is not solvable by radicals \cite{Bajaj}.

Nevertheless, for some special configurations, such a solution exists. The following theorem generalizes the Fagnano's result corresponding to the equal-weighted configurations ($\{m_j=1\}_{j=1}^4 $).

\begin{theorem} \label{Th3term42} Let Assumption 1 be fulfilled for the configuration $ \renewcommand{\arraystretch}{0.5} \left\{ \begin{array}{c|c|c|c} P_1 & P_2 & P_3 & P_4  \\ m_1 & m_2 & m_1 & m_2  \end{array} \right\} $. For any values of the weights $ m_1 $ and $ m_2 $, the position of the facility $ W $ providing the solution to the problem (\ref{Weber_uni}) is at the point of intersection  of
the quadrilateral $ P_1P_2P_3P_4 $ diagonals.
\end{theorem}

\textbf{Proof} consists in formal substitution of the coordinates
$$ x_{\ast}=\frac{(x_1-x_3)(x_2y_4-y_2x_4)-(x_2-x_4)(y_3x_1-y_1x_3)}{(x_3-x_1)(y_2-y_4)-(x_2-x_4)(y_3-y_1)}, $$ $$ y_{\ast}=\frac{(y_1-y_3)(x_2y_4-y_2x_4)-(y_2-y_4)(y_3x_1-y_1x_3)}{(x_3-x_1)(y_2-y_4)-(x_2-x_4)(y_3-y_1)}  $$
into the equations (\ref{StatPoint4}). \qed

An analytical approach is also effective for establishing the dynamics of the optimal facility location under variation of parameters. The following
result is a counterpart of Theorem \ref{Th3term31}.

\begin{theorem} \label{Th3term32} Let Assumption 1 be fulfilled. Let $ \{\mathfrak S_j\}_{j=1}^3 $ be defined by (\ref{S1S2S3}).  Set
$$
\mathfrak S_4=
\left| \begin{array}{rrr}
1 & 1 & 1 \\
x & x_3 & x_4 \\
y & y_3 & y_4
\end{array}
\right| \, .
$$
For any value of the weight $ m_3 $, the optimal facility $ W $ lies in the arc of the algebraic curve of the 12th degree  given by the equation
\begin{equation}
\frac{m_1^4 \mathfrak S_2^4}{|WP_1|^4} + \frac{m_2^4 \mathfrak S_1^4}{|WP_2|^4}+
\frac{m_4^4 \mathfrak S_4^4}{|WP_4|^4}-2\, \frac{m_1^2 m_2^2 \mathfrak S_1^2 \mathfrak S_2^2}{|WP_1|^2|WP_2|^2}
-2\, \frac{m_1^2 m_4^2 \mathfrak S_2^2 \mathfrak S_4^2}{|WP_1|^2|WP_4|^2}-
2\, \frac{m_2^2 m_4^2 \mathfrak S_1^2 \mathfrak S_4^2}{|WP_2|^2|WP_4|^2}=0 \, .
\label{4t1f_w}
\end{equation}
\end{theorem}

\textbf{Proof} is based on an idea similar to that used in the proof of Theorem \ref{Th3term31}, i.e. one should eliminate the variable $ m_3 $ from the system
(\ref{StatPoint4}) treated as a linear one with respect to the weights $ \{ m_j \}_{j=1}^4 $. \qed

\begin{example}\label{ex4t1f}
For the configuration
$$
\left\{\begin{array}{c|c|c|c}
P_1=(1,5) & P_2=(2,1) & P_3=(7,2) & P_4=(6,7) \\
m_1=3 & m_2=2 & m_3 & m_4=7/2
\end{array} %\mbox{ and }
\right\} \, ,
$$
find the locus of the facility $ W $ under variation of the weight $ m_3 $.
\end{example}

\textbf{Solution.} The complete expression (\ref{4t1f_w}) is rather cumbersome and we restrict  ourselves here with the presentation of its terms of the highest and the lowest degree
\begin{figure}[H]\center
{
\graphicspath{{Illustrations/}}
\includegraphics[scale=0.4]{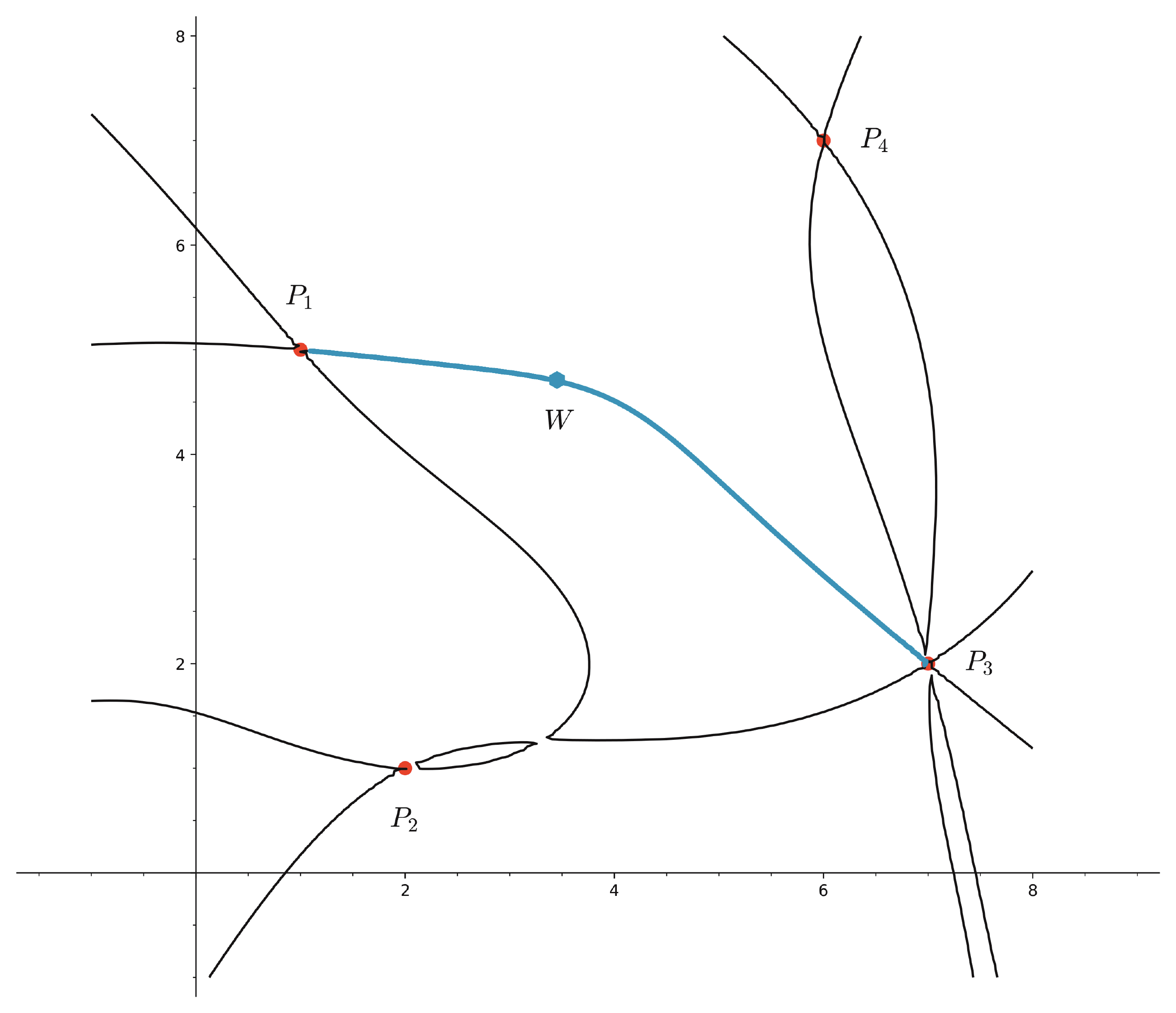}
   % \rule{4cm}{3cm}
    %\label{fig:curve_ed}
}
\caption{Example \ref{ex4t1f}. Optimal facility location under variation of the weight $ m_3 $} \label{weber_531}
\end{figure}
$$
49(7\,x+9\,y)(13\,x-9\,y)(3\,x-7\,y)(57\,x+23\,y)(x^2+y^2)^4+ \dots
$$
$$
-127\,164\,667\,059\,080\,x-53\,067\,633\,595\,480\,y+47\,008\,175\,916\,100=0 \, .
$$
The picture of the curve in the vicinity of the quadrilateral $ P_1P_2P_3P_4 $ is given in Fig. \ref{weber_531} with its only branch containing positions of the facility $ W $ displayed in blue boldface. The values of $ m_3 $ corresponding to this branch lie within the interval
\small
$$
\left[\frac{-696\sqrt{85}-952\sqrt{145}+\sqrt{-3621085+8696520\sqrt{493}}}{4930}, \frac{\sqrt{17069+1326\sqrt{130}}}{26} \right]
\approx [-0.834783,6.900360]
$$
\normalsize
with its ends corresponding to collision of the optimal facility with $ P_1 $ or $ P_3 $.
A sample point  $ W \approx (3.451796, 4.701666) $ marked in  Fig. \ref{weber_531}  matches the value
 $ m_3 \approx 1.394215 $.   \qed

%%%%%%%%%%%%%%%%%%%%%%%%%%%%%%%%%%%%%%%%%%%%%%%%%%%%%%
%%%%%%%%%%%%%%%%%%%%%%%%%%%%%%%%%%%%%%%%%%%%%%%%%%%%%%
%  4 TERMINALS
%%%%%%%%%%%%%%%%%%%%%%%%%%%%%%%%%%%%%%%%%%%%%%%%%%%%%%
%%%%%%%%%%%%%%%%%%%%%%%%%%%%%%%%%%%%%%%%%%%%%%%%%%%%%%

\section{Bifacility Case: Geometry}\label{SGeo}

First of all, we introduce the geometric observations given
by Georg Pick in the Mathematical Appendix of Weber's book \cite{Weber}.
We illustrate his algorithm with the following example.

\begin{example}\label{ft}
Find the optimal position for the facilities $ W_1 $ and $ W_2 $ for the problem (\ref{F_Weber}) where
$$
\left\{\begin{array}{c|c|c|c|}
P_1=(1,5) & P_2=(2,1) & P_3=(7,2) & P_4=(6,7) \\
m_1=3 & m_2=2 & m_3=3 & m_4=4
\end{array} %\mbox{ and }
 \ m=4  \right\} \, .
$$
\end{example}

\begin{figure}[H]\center
%\subfigure[Construction of two triangles on the sides $P_1P_2$ and $P_3P_4$]
{
\graphicspath{{Illustrations/}}
\includegraphics[scale=0.3]{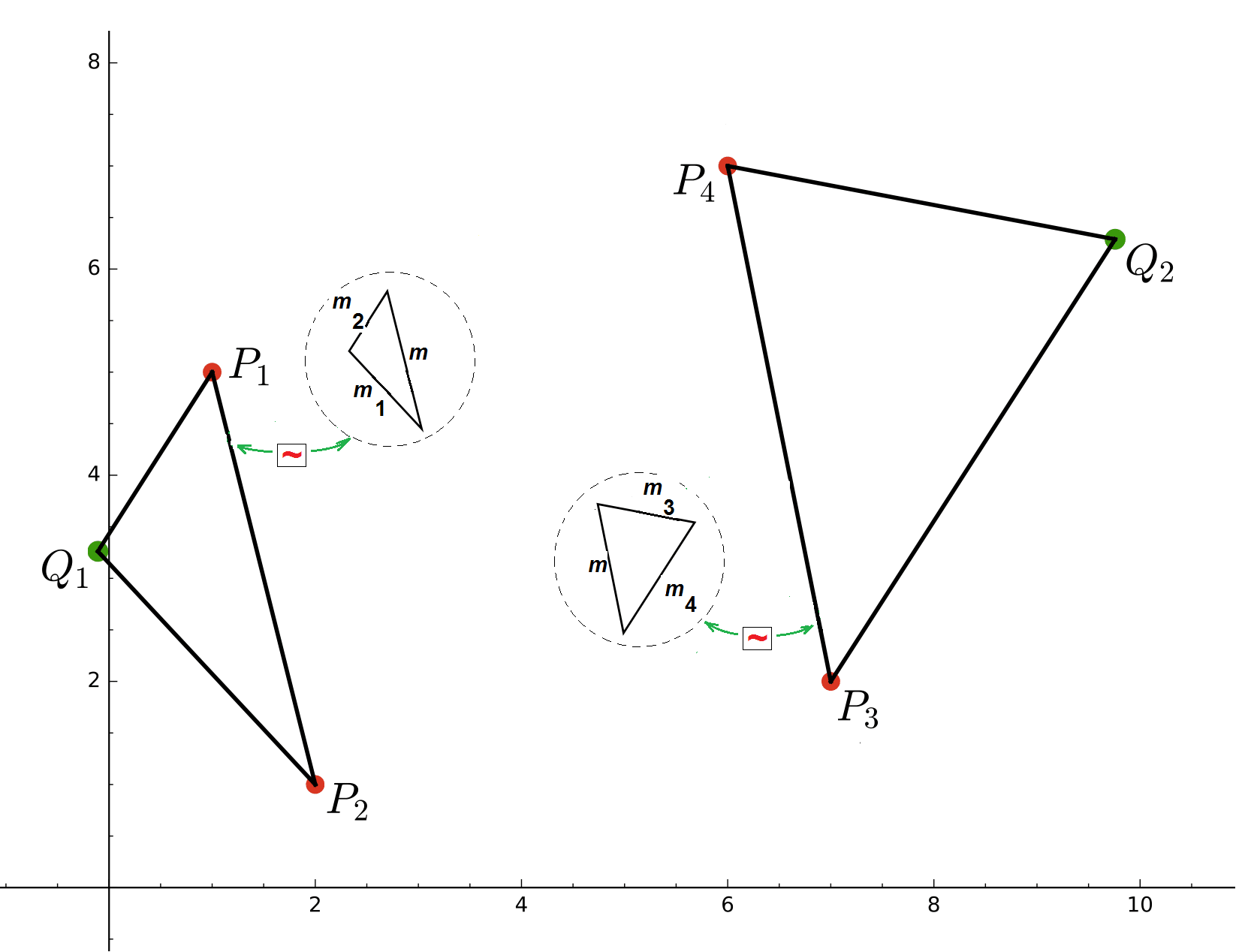}
    %\rule{4cm}{3cm}
    \label{fig:pick1}
    }
\caption{Example \ref{ft}. Construction of the points $ Q_1 $ and $ Q_2 $.}  \label{Pick_41}
\end{figure}

\begin{figure}[H]\center
{
\begin{minipage}[t]{145mm}
\begin{minipage}[t]{85mm}
\graphicspath{{Illustrations/}}
\includegraphics[scale=0.4]{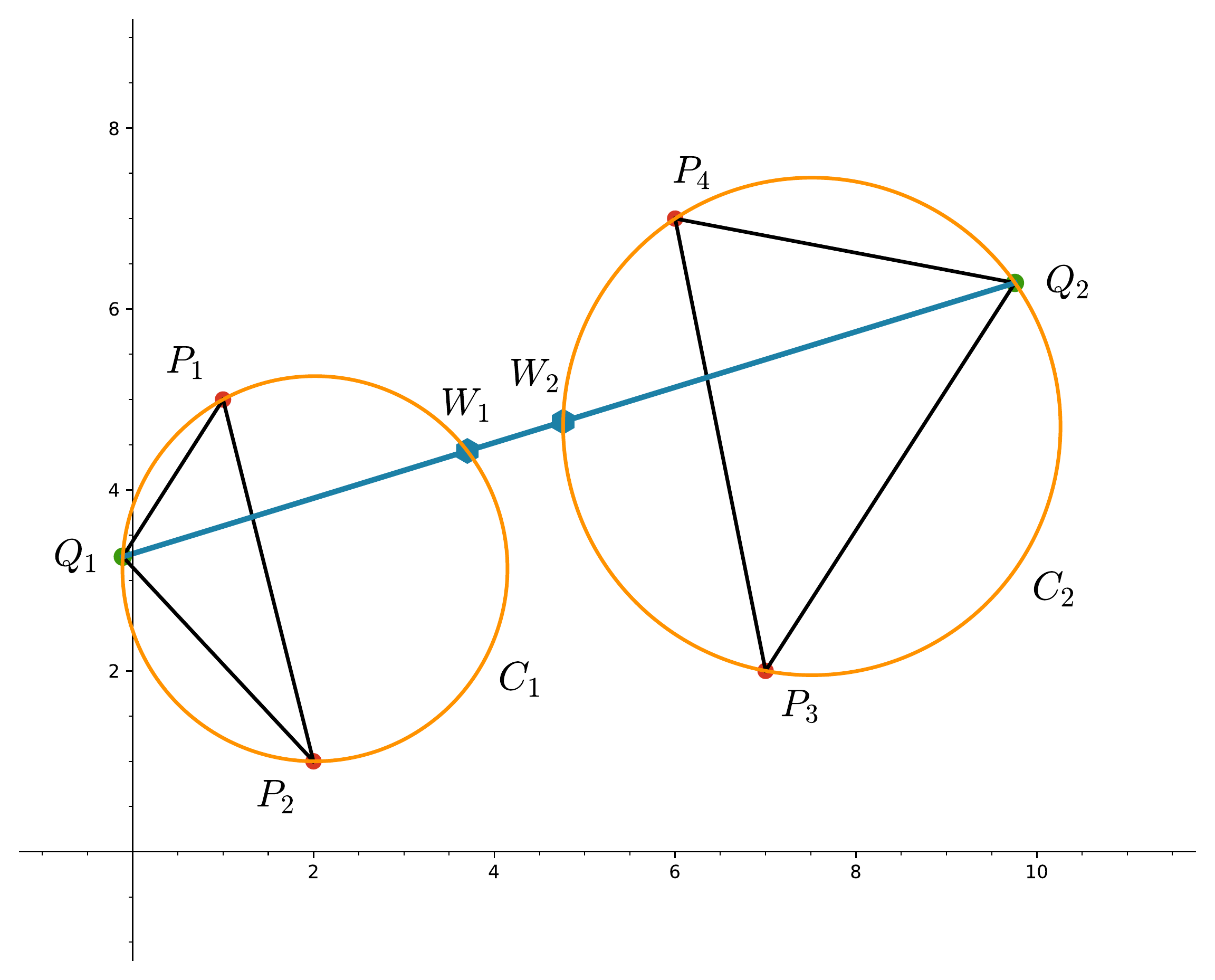}
\end{minipage}
\hfill
\begin{minipage}[t]{60mm}
\graphicspath{{Illustrations/}}
\includegraphics[scale=0.4]{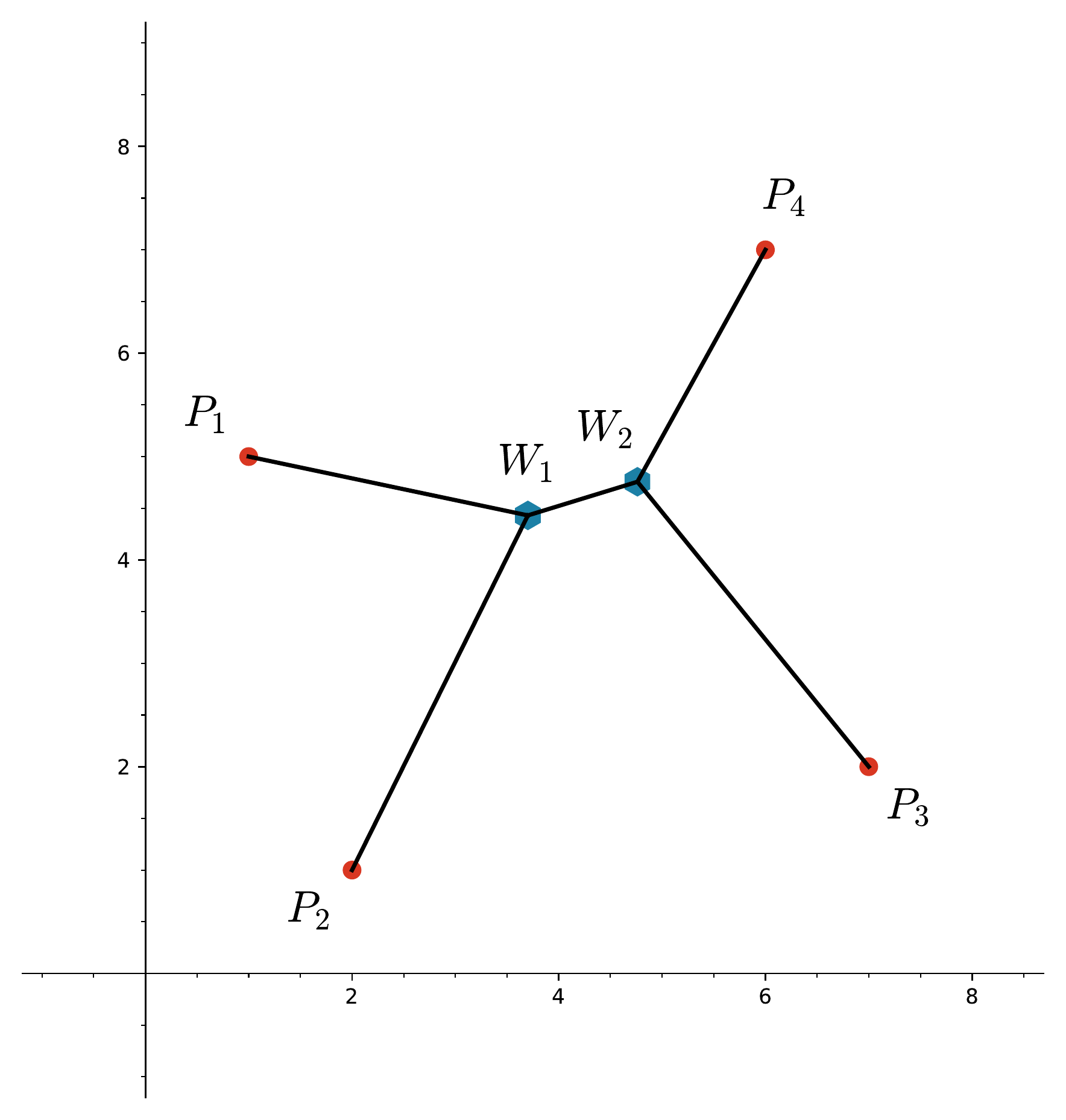}
\end{minipage}
\end{minipage}
}
\caption{Example \ref{ft}. Pick's construction of the Weber network}  \label{Pick_4}
\end{figure}

\textbf{Solution.} First find the point $ Q_1 $ lying on the opposite side of the line $ P_1P_2 $ with respect to the point $ P_3 $  and such that
\begin{equation}
 |P_1Q_1|=\frac{m_2}{m}|P_1P_2|, \ |P_2Q_1|=\frac{m_1}{m}|P_1P_2| \, .
 \label{similarity1}
\end{equation}
The exact coordinates of this point are given by (\ref{Q1}) where the substitution $ m_3 \to m $ is made.
Find then the second point $ Q_2 $ with the similar property with respect to the points $ P_3 $ and $ P_4 $ (Fig. \ref{Pick_41}):
$$ |P_3Q_2|=\frac{m_4}{m}|P_3P_4|, \ |P_4Q_2|=\frac{m_3}{m}|P_3P_4| \, . $$
Next, draw the circle $ C_1 $ circumscribing $P_1P_2Q_1$ and $ C_2 $ circumscribing $P_3P_4Q_2$.
%\begin{figure}[H]
%%\subfigure[Two circles circumscribing obtained triangles]
%{
%\graphicspath{{Illustrations/}}
%\includegraphics[scale=0.5]{pick2.pdf}
   % \rule{4cm}{3cm}
%   \label{fig:pick2}
%}
%\end{figure}
Finally draw the line through $ Q_1 $ and $ Q_2 $ (Fig. \ref{Pick_4}).

%\begin{figure}[H]\center
%{
%\graphicspath{{Illustrations/}}
%\includegraphics[scale=0.4]{pick3.pdf}
%   % \rule{4cm}{3cm}
%    \label{fig:pick3}
%}
%\end{figure}
\noindent The intersection points of this line with $ C_1 $ and $ C_2 $ are the position of the optimal facilities $ W_1 $ and $ W_2 $ for the network
with the corresponding (minimal) cost equal to $ m|Q_1Q_2| $.
\qed
%\begin{figure}[H]\center
%\subfigure[Weber points $W_1, W_2 $]
%{
%\graphicspath{{Illustrations/}}
%\includegraphics[scale=0.4]{pick4.pdf}
%   % \rule{4cm}{3cm}
%    \label{fig:pick4}
%}
%\caption{Pick's construction of the Weber network} \label{weber_53}
%\end{figure}
%

The suggested geometric construction just illustrated via an example, in general case  is subject to several extra assumptions. First of all, the point $ Q_1 $ exists and generates the triangle $P_1P_2Q_1$ iff the values of the weights
$m, m_1, m_2$ satisfy the restrictions
\begin{equation} \label{restrM1}
m<m_1+m_2,\  m_1<m+m_2, \ m_2<m+m_1 \, ,
\end{equation}
i.e. it is possible to construct a weight triangle $\{m , m_1, m_2\} $.
Similar restrictions are to be imposed onto the weights $m, m_3, m_4$
\begin{equation} \label{restrM2}
m<m_3+m_4,\  m_3<m+m_4, \ m_4<m+m_3 \, .
\end{equation}
The relations (\ref{similarity1}) then mean that the triangle $P_1P_2Q_1$ is \textit{similar} to the weight triangle $ \{m, m_1, m_2\}$.

Secondly, even if both weight triangles exist, the segment $ Q_1Q_2 $ might not cross either of the circles $ C_1 $ or $ C_2 $ or both in the points lying inside the quadrilateral $P_1P_2P_3P_4$.

\begin{example}\label{ftn}
For the configuration
$$
\left\{
\begin{array}{c|c|c|c|}
P_1=(2,4) & P_2=(1,1) & P_3=(6,2) & P_4=(5,5) \\
m_1=4 & m_2=2 & m_3=2 & m_4=5
\end{array} %\mbox{ and }
 \ m=4 \,
 \right\} ,
$$
\begin{figure}[H]\center
%\subfigure[Weber points $W_1, W_2 $]
{
\graphicspath{{Illustrations/}}
\includegraphics[scale=0.5]{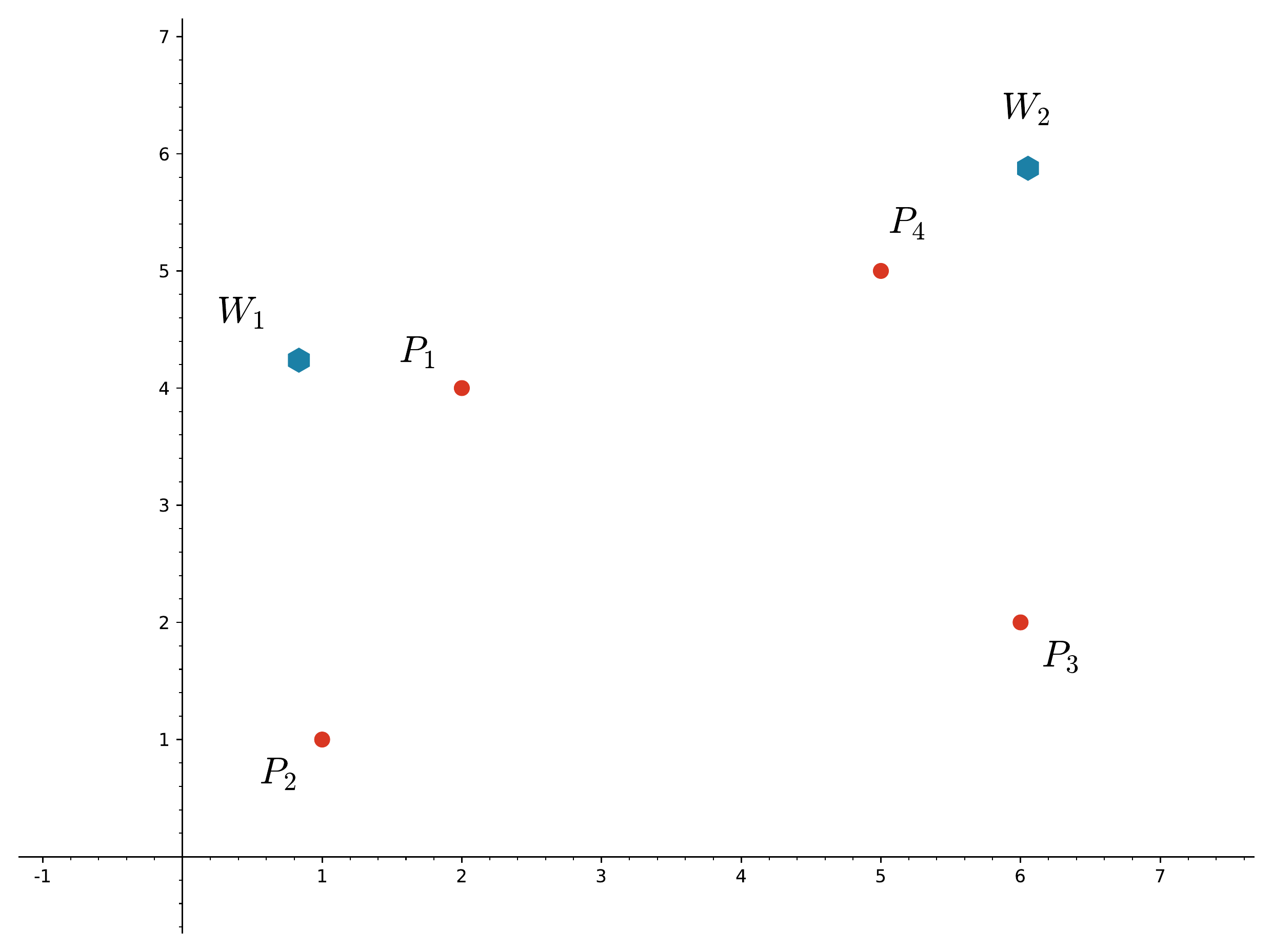}
   % \rule{4cm}{3cm}
    \label{fig:bad_config}
}
\caption{Example \ref{ftn}. Failure of the construction} \label{weber_532}
\end{figure}
\noindent one has $ Q_1\approx (0.42263, 4.10912),Q_2\approx (6.26863, 5.9437) $  with the points $ W_1\approx (0.83340, 4.23803) $ and $ W_2\approx (6.05325, 5.87612) $ lying outside the quadrilateral $ P_1P_2P_3P_4 $ (Fig. \ref{weber_532}).
\end{example}

Our next aim is now to establish the conditions for the feasibility of Pick's construction and to find the exact coordinates of the facilities.

\section{Bifacility Case: Analytics}\label{SAn}

\textbf{Assumption 2.} We will assume the weights of the problem to satisfy the restrictions (\ref{restrM1}) and
(\ref{restrM2}). From this follows that the values
\begin{eqnarray}
\mathbf k_{12}&=&(m+m_1+m_2)(m-m_1+m_2)(m+m_1-m_2)(-m+m_1+m_2), \label{k12} \\
\mathbf k_{34}&=&(m+m_3+m_4)(m-m_3+m_4)(m+m_3-m_4)(-m+m_3+m_4)\,   \label{k34}
\end{eqnarray}
are positive. Additionally we assume the fulfillment of the following inequalities:
\begin{eqnarray}
(m^2-m_1^2+m_2^2)/\sqrt{\mathbf k_{12}} + (m^2-m_4^2+m_3^2)/\sqrt{\mathbf k_{34}} & > & 0, \label{k12g0} \\
(m^2+m_1^2-m_2^2)/\sqrt{\mathbf k_{12}} + (m^2+m_4^2-m_3^2)/\sqrt{\mathbf k_{34}} & > & 0 \label{k34g0} \, .
\end{eqnarray}

The geometric sense of the latter restrictions will be clarified below.

\begin{theorem} \label{teo1} Let Assumptions 1 and 2 be fulfilled. Set
\begin{multline*}
\tau_1=\sqrt{{\mathbf k_{12}}} \Big[\sqrt{{\mathbf k_{34}}} (x_4-x_3)-(m^2+m_3^2-m_4^2)\, y_3-(m^2-m_3^2+m_4^2)\, y_4\Big]\\+2\, m^2 \sqrt{{\mathbf k_{12}}}\, y_2+{\mathbf k_{12}}
   (x_1-x_2)+(m^2+m_1^2-m_2^2)\Big[\sqrt{{\mathbf k_{34}}} (y_3-y_4) \\ +(m^2+m_1^2-m_2^2)\, x_1 +(m^2-m_1^2+m_2^2)
  \, x_2-(m^2+m_3^2-m_4^2)\, x_3-(m^2-m_3^2+m_4^2)\, x_4\Big],
\end{multline*}
\begin{multline*}
\tau_2=-\sqrt{{\mathbf k_{12}}} \Big[\sqrt{{\mathbf k_{34}}} (x_4-x_3)-(m^2+m_3^2-m_4^2)\, y_3-(m^2-m_3^2+m_4^2) y_4\Big]\\ -2\, m^2\, \sqrt{{\mathbf k_{12}}}\, y_1-{\mathbf k_{12}}
   (x_1-x_2)+(m^2-m_1^2+m_2^2)\Big[\sqrt{{\mathbf k_{34}}} (y_3-y_4) \\ +(m^2+m_1^2-m_2^2)\, x_1+(m^2-m_1^2+m_2^2)\,
   x_2-(m^2+m_3^2-m_4^2)\, x_3-(m^2-m_3^2+m_4^2) \,x_4\Big],
\end{multline*}
$$
\eta_1=\frac{1}{\sqrt{\mathbf k_{12}}}\left[ (m^2-m_1^2-m_2^2) \tau_1-2\, m_1^2 \tau_2 \right],\ \eta_2=\frac{1}{\sqrt{\mathbf k_{12}}}\left[2\, m_2^2 \tau_1 - (m^2-m_1^2-m_2^2) \tau_2  \right]
$$
and set the values for $ \tau_3, \tau_4, \eta_3, \eta_4 $ via the formulae obtained by the cyclic substitution for subscripts
$$
\left(
\begin{array}{cccc}
1 & 2 & 3 & 4 \\
3 & 4 & 1 & 2
\end{array}
\right)
\label{substitut}
$$
in the above expressions for $ \tau_1, \tau_2, \eta_1, \eta_2 $ correspondingly.

If all the values
\begin{eqnarray}
\delta_1&=&\eta_2 \,(x_1-x_2)+\tau_2\, (y_2-y_1),\
\delta_2=\eta_1\, (x_1-x_2)+\tau_1 \,(y_2-y_1),
\label{del12} \\
\delta_3&=&\eta_4\, (x_3-x_4)+\tau_4\, (y_4-y_3),\ \delta_4=\eta_3\, (x_3-x_4)+\tau_3\,
(y_4-y_3)
\label{del34}
\end{eqnarray}
and
\begin{equation}
\delta=-\frac{\delta _1 \left(m^2+m_1^2-m_2^2\right)}{\sqrt{\mathbf k_{12}}}-\frac{\delta _3 \left(m^2+m_3^2-m_4^2\right)}{\sqrt{\mathbf k_{34}}} +\left(\eta_1+\eta_2\right)
   \left(y_1-y_3\right)+\left(\tau_1+\tau_2\right) \left(x_1-x_3\right)
   \label{del}
\end{equation}
are positive then there exists a pair of points $W_1$ and $W_2$ lying inside $P_1P_2P_3P_4$ that provides
the global minimum value for the function (\ref{F_Weber}). The coordinates of the optimal facility $W_1$ are as follows:
\begin{eqnarray}
\begin{aligned}
x_*&=&x_1-\frac{2\, \delta _1\, m^2 \,\tau_1}{\sqrt{\mathbf k_{12}}\big[(\eta_1+\eta_2){}^2+(\tau_1+\tau_2){}^2\big]},
\end{aligned}
\label{W1x}
\end{eqnarray}
\begin{eqnarray}
\begin{aligned}
y_*&=&y_1-\frac{2\, \delta _1\, m^2\, \eta_1}{\sqrt{\mathbf k_{12}}\big[(\eta_1+\eta_2){}^2+(\tau_1+\tau_2){}^2\big]},
\end{aligned}
\label{W1y}
\end{eqnarray}
while those of $W_2$:
\begin{eqnarray}
\begin{aligned}
x_{**}&=&x_3-\frac{2 \, \delta _3\, m^2\, \tau_3}{\sqrt{\mathbf k_{34}}\big[(\eta_1+\eta_2){}^2+(\tau_1+\tau_2){}^2\big]},
\end{aligned}
\label{W2x}
\end{eqnarray}
\begin{eqnarray}
\begin{aligned}
y_{**}&=&y_3-\frac{2 \, \delta _3\, m^2\, \eta_3 }{\sqrt{\mathbf k_{34}}\big[(\eta_1+\eta_2){}^2+(\tau_1+\tau_2){}^2\big]}.
\end{aligned}
\label{W2y}
\end{eqnarray}
The corresponding minimum value for the function (\ref{F_Weber}) (i.e. the cost of the optimal network) equals
\begin{equation}
\mathfrak{C}=\frac{\sqrt{\left(\eta_1+\eta_2\right){}^2+\left(\tau_1+\tau_2\right){}^2}}{4 m^3} \, .
  \label{cost}
\end{equation}
\end{theorem}

%%%%%%%%%%%%%%%%%%%%%%%%%%%%%%%%%%%%%%%%%%%%%%%%%%%%%%
%  PROOF
%%%%%%%%%%%%%%%%%%%%%%%%%%%%%%%%%%%%%%%%%%%%%%%%%%%%%%
\textbf{Proof.} For brevity, we will use the following notation for the expression that appears nearly in any deduction of the proof:
\begin{equation}
\Delta=\left[(\eta_1+\eta_2)^2+(\tau_1+\tau_2)^2 \right] \, .
\label{DDelta}
\end{equation}
\textbf{(I)} We first present some directly verified relations between the values $ \tau $-s , $ \eta $-s and $ \delta $-s.
\begin{eqnarray}
\tau_1 &= & \frac{1}{2\, m^2} \left[\sqrt{\mathbf k_{12}}(\eta_1+\eta_2)+(m^2+m_1^2-m_2^2)(\tau_1+\tau_2) \right], \label{tau11} \\
\tau_2 &= & \frac{1}{2\, m^2} \left[-\sqrt{\mathbf k_{12}}(\eta_1+\eta_2)+(m^2-m_1^2+m_2^2)(\tau_1+\tau_2) \right], \nonumber \\
\tau_3 &= & \frac{1}{2\, m^2} \left[-\sqrt{\mathbf k_{34}}(\eta_1+\eta_2)-(m^2+m_3^2-m_4^2)(\tau_1+\tau_2) \right], \label{tau31} \\
\tau_4 &= & \frac{1}{2\, m^2} \left[\sqrt{\mathbf k_{34}}(\eta_1+\eta_2)-(m^2-m_3^2+m_4^2)(\tau_1+\tau_2) \right], \nonumber
\end{eqnarray}
\begin{equation}
\tau_1+\tau_2+\tau_3+\tau_4=0,\ \eta_1+\eta_2+\eta_3+\eta_4=0 \, ,
\label{tausEtas}
\end{equation}
\begin{equation}
\sum_{j=1}^4 ( x_j \tau_j + y_j \eta_j) = \frac{\Delta}{4\, m^4}  \, ;
\label{tausEtas0}
\end{equation}
\begin{eqnarray}
\tau_1^2+\eta_1^2 & = & \frac{m_1^2}{m^2} \Delta \, , \label{t1t1n2n} \\
\tau_1 \eta_2 - \tau_2 \eta_1 &= &\frac{\sqrt{\mathbf k_{12}}}{2m^2} \Delta, \label{t1n2}  \\
\tau_2 \eta_3 - \tau_3 \eta_2 & =& \frac{\sqrt{\mathbf k_{12} \mathbf k_{34}}}{4m^4}\left[ \frac{m^2-m_1^2+m_2^2}{\sqrt{\mathbf k_{12}}} + \frac{m^2-m_4^2+m_3^2}{\sqrt{\mathbf k_{34}}} \right] \Delta, \label{t2n3} \,
\end{eqnarray}
\begin{equation}
\delta_1+\delta_3=(x_1-x_3)(\eta_1+\eta_2)-(y_1-y_3)(\tau_1+\tau_2) \, , \label{delta13}
\end{equation}
\begin{eqnarray}
2\delta_2m_2^2&=&(m^2-m_1^2-m_2^2) \delta_1 -\sqrt{\mathbf k_{12}} \left[    (y_1-y_2)\eta_2 +(x_1-x_2)\tau_2 \right] \,  , \label{delta2m2} \\
2\delta_4m_4^2&=&(m^2-m_3^2-m_4^2) \delta_3 -\sqrt{\mathbf k_{34}} \left[    (y_3-y_4)\eta_4 +(x_3-x_4)\tau_4 \right] \,  . \label{delta4m4}
\end{eqnarray}

\textbf{(II)}  Consider the system of equations for determining stationary points of the objective function \eqref{F_Weber}:
\begin{eqnarray}
\frac{\partial F}{\partial x_{\ast}}&=&m_1 \, \frac{x_{\ast}-x_1}{|W_1P_1|}+m_2  \frac{x_{\ast}-x_2}{|W_1P_2|}+m \, \frac{x_{\ast}-x_{\ast \ast}}{|W_1W_2|}=0,
\label{grad_42} \\
\frac{\partial F}{\partial y_{\ast}}&=&m_1 \, \frac{y_{\ast}-y_1}{|W_1P_1|}+m_2  \,\frac{y_{\ast}-y_2}{|W_1P_2|}+m \, \frac{y_{\ast}-y_{\ast \ast}}{|W_1W_2|}=0, \label{grad_422} \\
\frac{\partial F}{\partial x_{\ast \ast}}&=&m_3 \, \frac{x_{\ast \ast}-x_3}{|W_2P_3|}+m_4 \, \frac{x_{\ast \ast}-x_4}{|W_2P_4|}+m \, \frac{x_{\ast \ast}-x_{\ast}}{|W_2W_1|}=0, \label{grad_423} \\
\frac{\partial F}{\partial y_{\ast \ast}}&=&m_3 \, \frac{y_{\ast \ast}-y_3}{|W_2P_3|}+m_4 \, \frac{y_{\ast \ast}-y_4}{|W_2P_4|}+m \, \frac{y_{\ast \ast}-y_{\ast}}{|W_2W_1|}=0\, . \label{grad_424}
\end{eqnarray}
Let us verify the validity of (\ref{grad_42}).
First establish the alternative representations for the coordinates \eqref{W1x} and \eqref{W1y}:
\begin{eqnarray}
\begin{aligned}
x_{\ast}&=x_2-\frac{2 \,m ^2\, \delta_2\, \tau_2}{\sqrt{\mathbf k_{12}}\Delta},
\label{xy_alt0}
\end{aligned}
\end{eqnarray}
\begin{eqnarray}
\begin{aligned}
y_{\ast}&=y_2-\frac{2\, m ^2\, \delta_2\, \eta_2}{\sqrt{\mathbf k_{12}}\Delta}.
\label{xy_alt}
\end{aligned}
\end{eqnarray}
Indeed, the difference of the right-hand sides of (\ref{W1x}) and (\ref{xy_alt0}) equals
\begin{eqnarray*}
x_1-x_2-\frac{2\, m ^2\, (\delta_1 \tau_1-\delta_2 \tau_2)}{\sqrt{\mathbf k_{12}}\Delta}
\end{eqnarray*}
and the numerator of the involved fraction can be transformed into
\begin{eqnarray*}
& \stackrel{\eqref{del12}}= &2\, m ^2\, \Big[\tau_1\eta_2(x_1-x_2)+\tau_1 \tau_2 (y_2-y_1)-\tau_2\eta_1(x_1-x_2)-\tau_2 \tau_1 (y_2-y_1)\Big]\\
&=& 2\, m^2\, (x_1-x_2) (\tau_1\eta_2-\tau_2\eta_1)
\stackrel{\eqref{t1n2}}
=(x_1-x_2)\sqrt{\mathbf k_{12}}\Delta\, .
\end{eqnarray*}
The equivalence of \eqref{xy_alt} and (\ref{W1y}) can be demonstrated in a similar manner.
Now express the segment lengths:
\begin{equation}
|W_1P_1|=\sqrt{(x_1-x_{\ast})^2+(y_1-y_{\ast})^2}
\stackrel{\eqref{W1x},\eqref{W1y}}= \frac{2 \delta_1 m^2}{\sqrt{\mathbf k_{12}}\Delta}\sqrt{\tau_1^2+\eta_1^2}
\stackrel{\eqref{t1t1n2n}}=
\frac{2 \, m \, m_1}{\sqrt{\mathbf k_{12}\mathstrut}} \, \frac{\delta_1}{ \sqrt{\Delta}} \,
\label{W1P1}
\end{equation}
and, similarly,
\begin{equation}
|W_1P_2|\stackrel{\eqref{xy_alt0},\eqref{xy_alt}}=\frac{2 \, m \, m_2}{\sqrt{\mathbf k_{12}\mathstrut}} \, \frac{\delta_2}{ \sqrt{\Delta}}.
\label{W1P2}
\end{equation}
With the aid of relations (\ref{W1x}), (\ref{xy_alt0}), (\ref{W1P1}) and (\ref{W1P2})
one can represent the first two terms in the left-hand side of the equality \eqref{grad_42} as
\begin{equation}
m_1 \, \frac{x_{\ast}-x_1}{|W_1P_1|} + m_2 \, \frac{x_{\ast}-x_2}{|W_1P_2|}=-\frac{m}{\sqrt{\Delta}}(\tau_1 + \tau_2)
\label{2sum}.
\end{equation}
The third summand in the equality \eqref{grad_42} needs more laborious manipulations. We first transform its numerator:
$$
x_{\ast}-x_{\ast \ast}\stackrel{\eqref{W1x},\eqref{W2x}}=x_1-x_3+\frac{2\, m^2}{\Delta}\, \Bigg[\frac{\delta_3\, \tau_3}{ \sqrt{\mathbf k_{34}}} -\frac{\delta_1\, \tau_1} {\sqrt{\mathbf k_{12}} }\Bigg].
$$
Now write down the following modification:
$$
 2\, m^2
\Bigg[\frac{\delta_3\, \tau_3}{ \sqrt{\mathbf k_{34}}} -\frac{\delta_1\, \tau_1} {\sqrt{\mathbf k_{12}} }\Bigg]
$$
$$
 \stackrel{\eqref{tau11},\eqref{tau31}} =
\left[ - (\eta_1+\eta_2)-\frac{m^2+m_3^2-m_4^2}{\sqrt{\mathbf k_{34}}} (\tau_1+\tau_2) \right]\delta_3 \\
- \left[(\eta_1+\eta_2)+\frac{m^2+m_1^2-m_2^2}{\sqrt{\mathbf k_{12}}} (\tau_1+\tau_2) \right]\delta_1
$$
$$
=-\left[ (\eta_1+\eta_2)(\delta_1+ \delta_3)+(\tau_1+\tau_2) \left\{  \frac{\delta _1 \left(m^2+m_1^2-m_2^2\right)}{\sqrt{\mathbf k_{12}}}+\frac{\delta _3 \left(m^2+m_3^2-m_4^2\right)}{\sqrt{\mathbf k_{34}}} \right\} \right]
$$
$$
\stackrel{\eqref{del}} =
-\left[ (\eta_1+\eta_2)(\delta_1+ \delta_3)+(\tau_1+\tau_2) \left\{ - \delta+ \left(\eta_1+\eta_2\right)
   \left(y_1-y_3\right)+\left(\tau_1+\tau_2\right) \left(x_1-x_3\right) \right\} \right]
$$
$$
=\delta (\tau_1+\tau_2) - (\eta_1+\eta_2)\left[  \delta_1+ \delta_3 +(\tau_1+\tau_2)(y_1-y_3) -(\eta_1+\eta_2)(x_1-x_3) \right] - \Delta (x_1-x_3)
$$
$$
\stackrel{\eqref{delta13}} = \delta (\tau_1+\tau_2) - \Delta (x_1-x_3).
$$
Finally,
\begin{equation}
x_{\ast}-x_{\ast \ast}=x_1-x_3+\frac{ \delta (\tau_1+\tau_2)\, - \Delta (x_1-x_3)} {\Delta}
= \frac{  \delta\, (\tau_1+\tau_2)} {\Delta}\, .
\label{xmx}
\end{equation}
Similarly the following equality can be deduced:
\begin{equation}
y_{\ast}-y_{\ast \ast}=
\frac{  \delta\, (\eta_1+\eta_2)} {\Delta},
\label{ymy}
\end{equation}
and both formulae yield
\begin{equation}
|W_1W_2|=\sqrt{(x_{\ast}-x_{\ast \ast})^2+(y_{\ast}-y_{\ast \ast})^2}=\frac{\delta}{\sqrt{\Delta}}\, .
\label{W1W2}
\end{equation}
Therefore, the last summand of equality (\ref{grad_42}) takes the form
$$
m \, \frac{x_{\ast}-x_{\ast \ast}}{|W_1W_2|}=m \, \frac{\delta \, (\tau_1+\tau_2) \sqrt{\Delta}}{\delta \Delta} =
m \, \frac{\tau_1+\tau_2}{\sqrt{\Delta}}.
$$
Summation this with (\ref{2sum}) yields $ 0 $ and this completes the proof of \eqref{grad_42}.

The validity of the remaining equalities (\ref{grad_422})--(\ref{grad_424}) can be established  in a similar way.

\vspace{1em}
\textbf{(III)} We now deduce the formula \eqref{cost} for the network cost. With the aid of the formulae (\ref{W1P1}), (\ref{W1P2}), (\ref{W1W2}) and their counterparts for the segment lengths $ |W_2P_3| $ and $ |W_2P_4| $, one gets
$$
m_1|W_1P_1|+m_2|W_1P_2| +m_3|W_2P_3|+m_4|W_2P_4|+ m |W_1W_2|
$$
$$
= \frac{2m}{\sqrt{\Delta}} \left( \frac{m_1^2 \,\delta_1}{\sqrt{\mathbf k_{12}}}+\frac{m_2^2 \,\delta_2}{\sqrt{\mathbf k_{12}}}+\frac{m_3^2\, \delta_3}{\sqrt{\mathbf k_{34}}}+\frac{m_4^2\, \delta_4}{\sqrt{\mathbf k_{34}}} +\frac{\delta}{2} \right)
$$
$$
\stackrel{\eqref{del}} =  \frac{2m}{\sqrt{\Delta}} \Bigg\{
\frac{\delta_1}{2\sqrt{\mathbf k_{12}}}(-m^2+m_2^2+m_1^2) +
\frac{\delta_3}{2\sqrt{\mathbf k_{34}}}(-m^2+m_3^2+m_4^2)+  \frac{m_2^2 \,\delta_2}{\sqrt{\mathbf k_{12}}} + \frac{m_4^2\, \delta_4}{\sqrt{\mathbf k_{34}}}
$$
$$
+\frac{1}{2}\left(\eta_1+\eta_2\right)
   \left(y_1-y_3\right)+\frac{1}{2} \left(\tau_1+\tau_2\right) \left(x_1-x_3\right) \Bigg\}
$$
$$
\stackrel{\eqref{delta2m2},\eqref{delta4m4}}=
\frac{2m}{\sqrt{\Delta}} \Bigg\{ -\frac{1}{2} (y_1-y_2)\eta_2 -\frac{1}{2} (x_1-x_2)\tau_2 -\frac{1}{2} (y_3-y_4)\eta_4 - \frac{1}{2}(x_3-x_4)\tau_4
$$
$$
+\frac{1}{2}\left(\eta_1+\eta_2\right)
   \left(y_1-y_3\right)+\frac{1}{2} \left(\tau_1+\tau_2\right) \left(x_1-x_3\right) \Bigg\}
$$
$$
=\frac{m}{\sqrt{\Delta}} \left\{y_1\eta_1+y_2 \eta_2 + y_4 \eta_4 + x_1 \tau_1+x_2\tau_2 + x_4 \tau_4
-x_3(\tau_1+\tau_2+\tau_4)-y_3(\eta_1+\eta_2+\eta_4)
\right\}
$$
$$
\stackrel{\eqref{tausEtas}}= \frac{m}{\sqrt{\Delta}} \sum_{j=1}^4 ( x_j \tau_j + y_j \eta_j) \stackrel{\eqref{tausEtas0}} =\frac{\sqrt{\Delta}}{4\, m^3} \, .
$$

\vspace{1em}
\textbf{(IV)} If the facilities $ W_1 $ and $ W_2 $ provide the solution to the problem (\ref{F_Weber}), they should lie inside the quadrilateral $ P_1P_2P_3P_4 $ \cite{Francis&Cabot}. Let us verify this condition checking the triangles $ P_1P_2W_1$, $ P_2W_2W_1 $ and $ P_2P_3W_1 $ are oriented counter\-clock\-wise. Indeed,
$$
\left|
\setlength{\arraycolsep}{3pt}
\begin{array}{ccc}
1 & 1  & 1 \\
x_1 & x_2 & x_{\ast} \\
y_1 & y_2 & y_{\ast}
\end{array}
\right|=
\left|
\setlength{\arraycolsep}{2pt}
\begin{array}{ccc}
1 & 0  & 0 \\
x_1 & x_2-x_1 & x_{\ast}-x_1 \\
y_1 & y_2-y_1 & y_{\ast}-y_1
\end{array}
\right|
\stackrel{\eqref{W1x},\eqref{W1y}}=-\frac{2\, \delta_1 m^2}{\sqrt{\mathbf k_{12}}\Delta}\left[(x_2-x_1)\eta_1-(y_2-y_1)\tau_1\right]
\stackrel{\eqref{del12}}
=\frac{2\,\delta_1 \delta_2 m^2}{\sqrt{\mathbf k_{12}}\Delta}\ ,
$$
\begin{equation}
\left|\begin{array}{ccc}
1 & 1  & 1 \\
x_2 & x_{\ast \ast} & x_{\ast} \\
y_2 & y_{\ast \ast} & y_{\ast}
\end{array}
\right|=
\left|\begin{array}{ccc}
1 & 0  & 0 \\
x_2 & x_{\ast \ast} - x_{\ast} & x_{\ast}-x_2 \\
y_2 & y_{\ast \ast} - y_{\ast} & y_{\ast}-y_2
\end{array}
\right|
\label{det1}
\end{equation}
$$
\stackrel{\eqref{xmx},\eqref{ymy}}=\frac{2\, \delta \delta_2 m^2}{\sqrt{\mathbf k_{12}}\Delta^2}
\left|\begin{array}{cc}
\tau_1+\tau_2 & \tau_2 \\
\eta_1 + \eta_2 & \eta_2
\end{array}
\right|=
\frac{2\, \delta \delta_2 m^2}{\sqrt{\mathbf k_{12}}\Delta^2}(\tau_1 \eta_2-\tau_2 \eta_1)
\stackrel{\eqref{t1n2}}=
\frac{2\, \delta \delta_2 m^2}{\sqrt{\mathbf k_{12}}\Delta^2}\frac{\sqrt{\mathbf k_{12}}\Delta}{2m^2}=
\frac{\delta \delta_2}{\Delta} \, .
$$
Due to assumptions of positivity of all the deltas, both determinants are positive. In order to prove positivity of the determinant
\begin{equation}
\left|\begin{array}{ccc}
1 & 1  & 1 \\
x_2 & x_{3} & x_{\ast} \\
y_2 & y_{3} & y_{\ast}
\end{array}
\right|
\label{det2}
\end{equation}
let us extract it from the alternative computation of the determinant (\ref{det1}).
$$
\left|\begin{array}{ccc}
1 & 1  & 1 \\
x_2 & x_{\ast \ast} & x_{\ast} \\
y_2 & y_{\ast \ast} & y_{\ast}
\end{array}
\right|
=
\left|\begin{array}{ccc}
1 & 1  & 1 \\
x_2 & x_3 & x_{\ast} \\
y_2 & y_3 & y_{\ast}
\end{array}
\right|+
\left|\begin{array}{ccc}
1 &    0                   & 0 \\
x_1 & x_{\ast \ast}-x_3    & x_{\ast } -x_2 \\
y_1 & y_{\ast \ast} - y_3 & y_{\ast} - y_2
\end{array}
\right| \, .
$$
Therefore, the determinant (\ref{det2}) equals
$$
\frac{\delta \delta_2}{\Delta}-\frac{2\, m^4 \delta_2 \delta_3}{\sqrt{\mathbf k_{12}\mathbf k_{34}} \Delta^2} (\tau_3\eta_2-\tau_2 \eta_3)
\stackrel{\eqref{t2n3}}=
\frac{\delta \delta_2}{\Delta}+\frac{\delta_2 \delta_3}{\Delta}\left[ \frac{m^2-m_1^2+m_2^2}{\sqrt{\mathbf k_{12}}} + \frac{m^2-m_4^2+m_3^2}{\sqrt{\mathbf k_{34}}} \right]
$$
and it is positive due to the assumption (\ref{k12g0}).

\vspace{1em}
\textbf{(V)} We finally prove that the formulae (\ref{W1x})--(\ref{W2y}) furnish the minimal value for the function (\ref{F_Weber}). For this aim, represent the Hessian of this function
\begin{equation}
\mathcal H (F)= \left[\begin{array}{cccc}
\partial^2 F /\partial x_{\ast}^2 & \partial^2 F / \partial x_{\ast} \partial y_{\ast} &
\partial^2 F / \partial x_{\ast} \partial x_{\ast \ast} & \partial^2 F / \partial x_{\ast} \partial y_{\ast \ast}  \\
\star & \partial^2 F /\partial y_{\ast}^2 & \partial^2 F / \partial y_{\ast} \partial x_{\ast \ast}  &
\partial^2 F / \partial y_{\ast} \partial y_{\ast \ast}  \\
\star & \star & \partial^2 F /\partial x_{\ast \ast}^2 & \partial^2 F /\partial x_{\ast \ast} \partial  y_{\ast \ast} \\
\star & \star & \star & \partial^2 F /\partial y_{\ast \ast}^2
\end{array}
\right]
\label{Hessian}
\end{equation}
as a product
$$
= \mathcal M \cdot \mathcal M^{\top}
$$
where
$$
\mathcal M =
\left[\begin{array}{rrrrr}
\frac{\sqrt{m_1} (y_{\ast}-y_1)}{|W_1P_1|^{3/2}} &
\frac{\sqrt{m_2} (y_{\ast}-y_2)}{|W_1P_2|^{3/2}} & \frac{\sqrt{m} (y_{\ast}-y_{\ast \ast})}{|W_1W_2|^{3/2}} &  0 \qquad & 0 \qquad \\
-\frac{\sqrt{m_1} (x_{\ast}-x_1)}{|W_1P_1|^{3/2}} &
-\frac{\sqrt{m_2} (x_{\ast}-x_2)}{|W_1P_2|^{3/2}} & -\frac{\sqrt{m} (x_{\ast}-x_{\ast \ast})}{|W_1W_2|^{3/2}} &  0 \qquad & 0 \qquad\\
0 \qquad  & 0 \qquad & \frac{\sqrt{m} (y_{\ast \ast}-y_{\ast})}{|W_1W_2|^{3/2}} & \frac{\sqrt{m_3} (y_{\ast \ast}-y_3)}{|W_2P_3|^{3/2}}  & \frac{\sqrt{m_4} (y_{\ast \ast}-y_4)}{|W_2P_4|^{3/2}} \\
0 \qquad  & 0 \qquad & -\frac{\sqrt{m} (x_{\ast \ast}-x_{\ast})}{|W_1W_2|^{3/2}} & -\frac{\sqrt{m_3} (x_{\ast \ast}-x_3)}{|W_2P_3|^{3/2}}  & -\frac{\sqrt{m_4} (x_{\ast \ast}-x_4)}{|W_2P_4|^{3/2}}
\end{array}
\right]_{4\times 5}
$$
and $ {}^{\top} $ stands for transposition. Therefore, Hessian (\ref{Hessian}) can be interpreted as the Gramian of the rows of the matrix $ \mathcal M $. The minor of the latter obtained by deleting the third its column equals
$$
\frac{\sqrt{m_1m_2m_3m_4}}{\left(|W_1P_1|\cdot |W_1P_2| \cdot |W_2P_3| \cdot |W_2P_4| \right)^{3/2}}
\left|\begin{array}{cc}
y_{\ast} - y_1 & y_{\ast} - y_2 \\
x_{\ast} - x_1 & x_{\ast} - x_2
\end{array}
\right|\cdot
\left|\begin{array}{cc}
y_{\ast \ast } - y_3 & y_{\ast \ast} - y_4 \\
x_{\ast \ast} - x_3 & x_{\ast \ast} - x_4
\end{array}
\right|
$$
$$
=
\frac{\sqrt{m_1m_2m_3m_4}}{\left(|W_1P_1|\cdot |W_1P_2| \cdot |W_2P_3| \cdot |W_2P_4| \right)^{3/2}}
\left|\begin{array}{ccc}
1 & 1  & 1 \\
x_1 & x_2 & x_{\ast} \\
y_1 & y_2 & y_{\ast}
\end{array}
\right|\cdot
\left|\begin{array}{ccc}
1 & 1  & 1 \\
x_3 & x_4 & x_{\ast \ast} \\
y_3 & y_4 & y_{\ast \ast}
\end{array}
\right|
$$
and, under the assumptions of the theorem, is nonzero for \emph{any choice of the points} $ W_1 $ and $ W_2 $ inside the quadrilateral $ P_1P_2P_3P_4 $. Consequently, the rank of the matrix $ \mathcal M  $ equals $ 4 $, its rows are linearly independent, and their Gramian is a positive definite matrix. From the Convex Optimization theory \cite{Saaty,Boyd_Optim}, it  follows that the function (\ref{F_Weber}) is strictly convex inside the \emph{convex} (due to Assumption 1) domain given as the Cartesian product  $ P_1P_2P_3P_4 \times P_1P_2P_3P_4  $. Therefore the solution of the system (\ref{grad_42}) -- (\ref{grad_424}) provides the global minimum value for this function. \qed

\begin{remark} In \cite{Cooper}, it is proved that the function
$$\widetilde F(W_1,W_2)= m_1|W_1P_1|+m_2|W_1P_2| +m_3|W_2P_3|+m_4|W_2P_4| $$
is neither convex nor concave if treated as a function of variables $ x_{\ast}, y_{\ast}, x_{\ast \ast}, y_{\ast \ast} $ \emph{and} $ \{m_j\}_{j=1}^4 $. This result should be distinguished from that claimed in the part \mbox{\textbf{(V)}} of the proof of Theorem \ref{teo1}: the objective function (\ref{F_Weber}) contains an extra term and the weights are not treated as variables.
\end{remark}

The result of Theorem \ref{teo1} claims that the bifacility Weber problem for four terminals is solvable by radicals, and thus we get a natural extension of the three-terminal problem solution given in Theorem \ref{Th3term1}. An additional correlation between these two results can be watched, namely that the denominators of all the formulae for the facilities coordinates contain the explicit expression for the cost of the corresponding network. It looks like every facility ``knows'' the cost of the network which this point is a part of.

%%%%%%%%%%%%%%%%%%%%%%%%%%%%%%%%%%%%%%%%%%%%%%%%%%%%%%
%  EXAMPLE 1
%%%%%%%%%%%%%%%%%%%%%%%%%%%%%%%%%%%%%%%%%%%%%%%%%%%%%%

%\begin{figure}[H]\center
%\graphicspath{{Illustrations/}}
%\includegraphics[scale=0.9]{wweber-4-2}
%\caption[]{The Weber network for four terminals under the configuration of weights from example \ref{ex1}}\label{wweber-4-2}
%\end{figure}

\begin{example}\label{ex1}
Find the exact coordinates of the facilities $ W_1 $ and $ W_2 $ for Example \ref{ft}.
\end{example}

\textbf{Solution.} The conditions of Theorem \ref{teo1} are fulfilled: the values $ \{\delta_j\}_{j=1}^4 $ and $ \delta $  are positive. Formulae (\ref{W1x})--(\ref{cost}) then give the
coordinates for the facilities
\begin{eqnarray*}
W_1 &=& \Bigg(\frac{\scriptstyle{2266800+772027 \sqrt{15}+453552 \sqrt{33}+246177 \sqrt{55}}}{\scriptstyle{48 \left(22049+2085 \sqrt{15}+945
   \sqrt{33}+2559 \sqrt{55}\right)}},
    \frac{\scriptstyle{1379951+201984 \sqrt{15}+97279 \sqrt{33}+154368 \sqrt{55}}}{\scriptstyle{16 \left(22049+2085 \sqrt{15}+945
   \sqrt{33}+2559 \sqrt{55}\right)}} \Bigg) \\
   & & \approx ( 3.701271,  4.430843) \, ; \\
W_2 &=& \Bigg(\frac{\scriptstyle{188467345+18613485\sqrt{15}+7149825\sqrt{33}+20949207\sqrt{55}}}{\scriptstyle{1760\left(22049+2085 \sqrt{15}+945
   \sqrt{33}+2559 \sqrt{55}\right)}},  \frac{\scriptstyle{188346565+19265895\sqrt{15}+20525157\sqrt{55}+7187445 \ \sqrt{11}}}
   {\scriptstyle{1760\left(22049+2085 \sqrt{15}+945
   \sqrt{33}+2559 \sqrt{55}\right)}} \Bigg)  \\
   & & \approx ( 4.761622, 4.756175)
\end{eqnarray*}
with the cost of the network
$$
\mathfrak C=\frac{1}{8}\sqrt{44098+4170\sqrt{15}+5118\sqrt{55}+1890\sqrt{33}}  \approx  41.280608.
$$
\qed

One can now verify directly correctness of Pick's geometric solution from Section \ref{SGeo}:

\begin{cor} Under the conditions of the theorem, the facilities $ W_1,W_2 $ and the points $ Q_1, Q_2 $ are collinear. The cost of the network equals $ m|Q_1Q_2| $.
\end{cor}

We outline briefly the meaning of the assumptions from Theorem \ref{teo1}. First, we
 concern the values (\ref{k12}) and  (\ref{k34}).   The values $\frac{1}{4} \sqrt{\mathbf k_{12}}$ and $\frac{1}{4} \sqrt{\mathbf k_{34}}$ equal the areas of the weight triangles introduced in Section \ref{SGeo}. Next, due to the law of cosines, one has
 $$  (m^2-m_1^2+m_2^2)/\sqrt{\mathbf k_{12}}= \cot \beta_1 , (m^2-m_4^2+m_3^2)/\sqrt{\mathbf k_{34}}=\cot \beta_4 $$
where $ \beta_1 $ and $ \beta_4 $ are the angles of the corresponding weight triangles
(Fig. \ref{fig:WeightQuadr} (b)). Therefore the condition (\ref{k12g0}) is equivalent to the fact that the angle of the \textbf{weight quadrilateral}, as illustrated in Fig. \ref{fig:WeightQuadr} (b), is less than $ \pi $. Together with the condition (\ref{k34g0}) this implies that the weight quadrilateral is convex. This condition is stated in
Theorem \ref{teo1} as a sufficient one for the existence of solution to the bifacility Weber problem. As yet we have failed to prove that it is stiff enough to be a necessary one.

%\begin{figure}[H]
%{
%\graphicspath{{Illustrations/}}
%\includegraphics[scale=0.5]{WeightQuadr.pdf}

%}
%\caption{Weight (a) triangle and (b) quadrangle }
%\end{figure}

Positivity of all the values (\ref{del12})-(\ref{del34})  guarantees the non-collision of the facilities $W_1 $ and $ W_2 $ with the terminals $ \{ P_j \}_{j=1}^4 $.
Finally, due to the equality (\ref{W1W2}), the condition (\ref{del}) guarantees the non-collision of the facilities $ W_1 $ and $ W_2 $, i.e. the non-degeneracy of the network with two assumed facilities. It is possible to deduce some alternative representation for this value, say more ``symmetric'' with respect to the involved parameters. For instance, the following equality
$$
\delta=\frac{1}{\left|\begin{array}{cc} x_1 - x_2 & y_1 - y_2 \\ x_3 - x_4 & y_3 - y_4 \end{array}  \right|}
\left|\begin{array}{cccr}
1 & x_1 & y_1 & - \delta_1 (m^2+m_1^2-m_2^2)/ \sqrt{\mathbf k_{12}} \\
1 & x_2 & y_2 & - \delta_2 (m^2-m_1^2+m_2^2)/ \sqrt{\mathbf k_{12}} \\
1 & x_3 & y_3 &  \delta_3 (m^2+m_3^2-m_4^2)/ \sqrt{\mathbf k_{34}} \\
1 & x_4 & y_4 &  \delta_4 (m^2-m_3^2+m_4^2)/ \sqrt{\mathbf k_{34}}
\end{array}
\right|
$$
is valid provided that the edges $ P_1P_2 $ and $ P_3P_4 $ are non-parallel.

The more detailed representation is as follows:
$$ \delta=4\, m^4 \left[(x_1-x_3)(x_2-x_4)+(y_1-y_3)(y_2-y_4) \right] $$
$$+\frac{2}{\sqrt{\mathbf k_{34}}}\left[(m_1^2-m_2^2) \mathbf k_{34} -m^2(m^2-m_1^2+m_2^2)(m^2-m_3^2-m_4^2) \right] \cdot
\left|
\begin{array}{ccc}
1&  1 & 1\\
x_2 & x_3 & x_4 \\
y_2 & y_3 & y_4
\end{array}
\right|
$$
$$+\frac{2}{\sqrt{\mathbf k_{34}}}\left[(m_2^2-m_1^2) \mathbf k_{34} -m^2(m^2+m_1^2-m_2^2)(m^2-m_3^2-m_4^2) \right] \cdot
\left|
\begin{array}{ccc}
1&  1 & 1\\
x_1 & x_3 & x_4 \\
y_1 & y_3 & y_4
\end{array}
\right|
$$
$$+\frac{2}{\sqrt{\mathbf k_{12}}}\left[(m_3^2-m_4^2) \mathbf k_{12} -m^2(m^2-m_3^2+m_4^2)(m^2-m_1^2-m_2^2) \right] \cdot
\left|
\begin{array}{ccc}
1&  1 & 1\\
x_1 & x_2 & x_4 \\
y_1 & y_2 & y_4
\end{array}
\right|
$$
$$+\frac{2}{\sqrt{\mathbf k_{12}}}\left[(m_4^2-m_3^2) \mathbf k_{12} -m^2(m^2+m_3^2-m_4^2)(m^2-m_1^2-m_2^2) \right] \cdot
\left|
\begin{array}{ccc}
1&  1 & 1\\
x_1 & x_2 & x_3 \\
y_1 & y_2 & y_3
\end{array}
\right|
$$
$$
+\frac{1}{\sqrt{\mathbf k_{12}\mathbf k_{34}}}\left[\left(\sqrt{\mathbf k_{12}}-\sqrt{\mathbf k_{34}}\right)^2m^4 -\left( \sqrt{\mathbf k_{34}}(m_2^2-m_1^2)+ \sqrt{\mathbf k_{12}}(m_3^2-m_4^2) \right)^2\right]
$$
$$
\times \left[ (x_4-x_3)(x_2-x_1)+(y_4-y_3)(y_2-y_1) \right] \, .
$$
This representation permits one to relate the general Weber problem to its important particular case:

\begin{cor}
For the equal weighted case $ \{m_j=1\}_{j=1}^4 $, $ m=1 $, the expression for $\delta$ can be represented in the form
\begin{equation}
 \delta=\frac{8}{\sqrt{3}}\left[x_3-x_1,y_3-y_1\right] \cdot \left[\begin{array}{rr} \sqrt{3}/2 & 1/2 \\ -1/2 &  \sqrt{3}/2 \end{array} \right] \cdot \left[\begin{array}{c} x_4-x_2 \\ y_4-y_2 \end{array} \right].
 \label{delta}
\end{equation}
This value is positive iff the angle between the diagonal $ \overrightarrow{P_1P_3} $ of the quadrilateral and the other diagonal $ \overrightarrow{P_2P_4} $ turned \allowbreak through by $ \pi/6 $ clockwise is acute. Equivalently, if we denote by $ \psi $ the angle between the diagonal vectors $ \overrightarrow{P_1P_3} $ and $ \overrightarrow{P_2P_4} $ then $ \delta $ is positive iff $ \psi < \pi/2 +\pi/6=2\, \pi/3 $. This confirms the known condition for the existence of a full Steiner tree for the terminals $ \{P_j\}_{j=1}^4 $, i.e. the points $ S_1 $ and $ S_2 $ providing the solution to the problem
$$
\min_{\{S_1,S_2\} \subset \mathbb R^2} \left\{
|P_1S_1|+|P_2S_1| +|P_3S_2|+|P_4S_2|+ |S_1S_2| \right\} \, .
$$
Formulae (\ref{W1x})--(\ref{W2y}) yield then the coordinates of these (Steiner) points  with the length of the minimal tree equal to
$$\mathfrak{C}=\frac{1}{2}\sqrt{A^2+B^2}$$
where
%$$\mathfrak{C}=\frac{1}{2}\sqrt{(\sqrt{3}(x_1-x_2-x_3+x_4)+(y_1+y_2-y_3-y_4))^2+((x_1+x_2-x_3-x_4)+\sqrt{3}(-y_1+y_2+y_3-y_4))^2}$$
$$ A= \sqrt{3}(x_1-x_2-x_3+x_4)+(y_1+y_2-y_3-y_4)\, ,
B=(x_1+x_2-x_3-x_4)+\sqrt{3}(-y_1+y_2+y_3-y_4)\, .
$$
%\begin{eqnarray*}
%A&=& \sqrt{3}(x_1-x_2-x_3+x_4)+(y_1+y_2-y_3-y_4), \\
%B&=&(x_1+x_2-x_3-x_4)+\sqrt{3}(-y_1+y_2+y_3-y_4)\, .
%\end{eqnarray*}
\end{cor}

%%%%%%%%%%%%%%%%%%%%%%%%%%%%%%%%%%%%%%%%%%%%%%%%%%%%%%
%  ANALYSIS
%%%%%%%%%%%%%%%%%%%%%%%%%%%%%%%%%%%%%%%%%%%%%%%%%%%%%%
\section{Solution Analysis}\label{SPar}

Though the analytical solution obtained in the previous section looks cumbersome in comparison with elegancy of the geometrical one described in Section \ref{SGeo}, it possesses two undeniable advantages over the latter.
First, it provides one with a unique opportunity to analyze the dynamics of the network under variation of the parameters of the configuration and to find the bifurcation values for these parameters, i.e. those responsible for the topology degeneracy. The second benefit is a wonderful occasion
for replacing the formal proofs of  some statements below (Theorems \ref{ThDyn1}, \ref{ThDyn2} and \ref{Th_cost}) with the words ``\dots via direct substitution of the  formulae (\ref{W1x})-(\ref{W2y})'' .

We first treat the case where the coordinates of a terminal are variated.
The following result is an evident counterpart of Theorem \ref{Th3term2}.

\begin{theorem} \label{Th4Inv} If the facilities $ W_1 $ and $ W_2 $ give the solution to the problem (\ref{F_Weber}) for some configuration $ \renewcommand{\arraystretch}{0.5} \left\{ \begin{array}{c|c|c|c} P_1 & P_2 & P_3 & P_4 \\ m_1 & m_2 & m_3 & m_4 \end{array} \right\} $  then these facilities remain unchanged for the  con\-fi\-gu\-ration \break $ \renewcommand{\arraystretch}{0.5}  \left\{ \begin{array}{c|c|c|c} P_1 & P_2 & \widetilde P_3 & P_4 \\ m_1 & m_2 & m_3 & m_4 \end{array} \right\}  $
with any position of the  terminal $ \widetilde P_3 $ in the half-line $ W_2P_3 $.
\end{theorem}

\begin{example}\label{ex2} For the configuration
 $$
\left\{\begin{array}{c|c|c|c|}
P_1=(1,5) & P_2=(2,1) & P_3 & P_4=(6,7) \\
m_1=3 & m_2=2 & m_3=3 & m_4=4
\end{array}\ m=4 %\mbox{ and }
\right\} \, ,
$$
find the loci of the facilities $ W_1 $ and $ W_2 $ under variation of the terminal $ P_3 $ moving somehow from the starting position
at $ (9,2) $ towards $ P_2 $.
\end{example}

\textbf{Solution.} It turns out that when $ P_3 $ wanders, the facility $ W_1 $ moves along the arc of the circle $ C_1 $ introduced in Theorem \ref{Th3term3} (with the replacement of $ m_3 $ by $ m $). It is given by the equality
\begin{equation}
 \left(x-\left\{\frac{3}{2}+\frac{2}{15}\sqrt{15}\right\}\right)^2+ \left(y-\left\{ 3+\frac{1}{30}\sqrt{15} \right\} \right)^2 = \frac{68}{15}.
 \label{C1_ex3}
\end{equation}

%FIG 2
\begin{figure}[H]\center
\graphicspath{{Illustrations/}}
\includegraphics[scale=0.7]{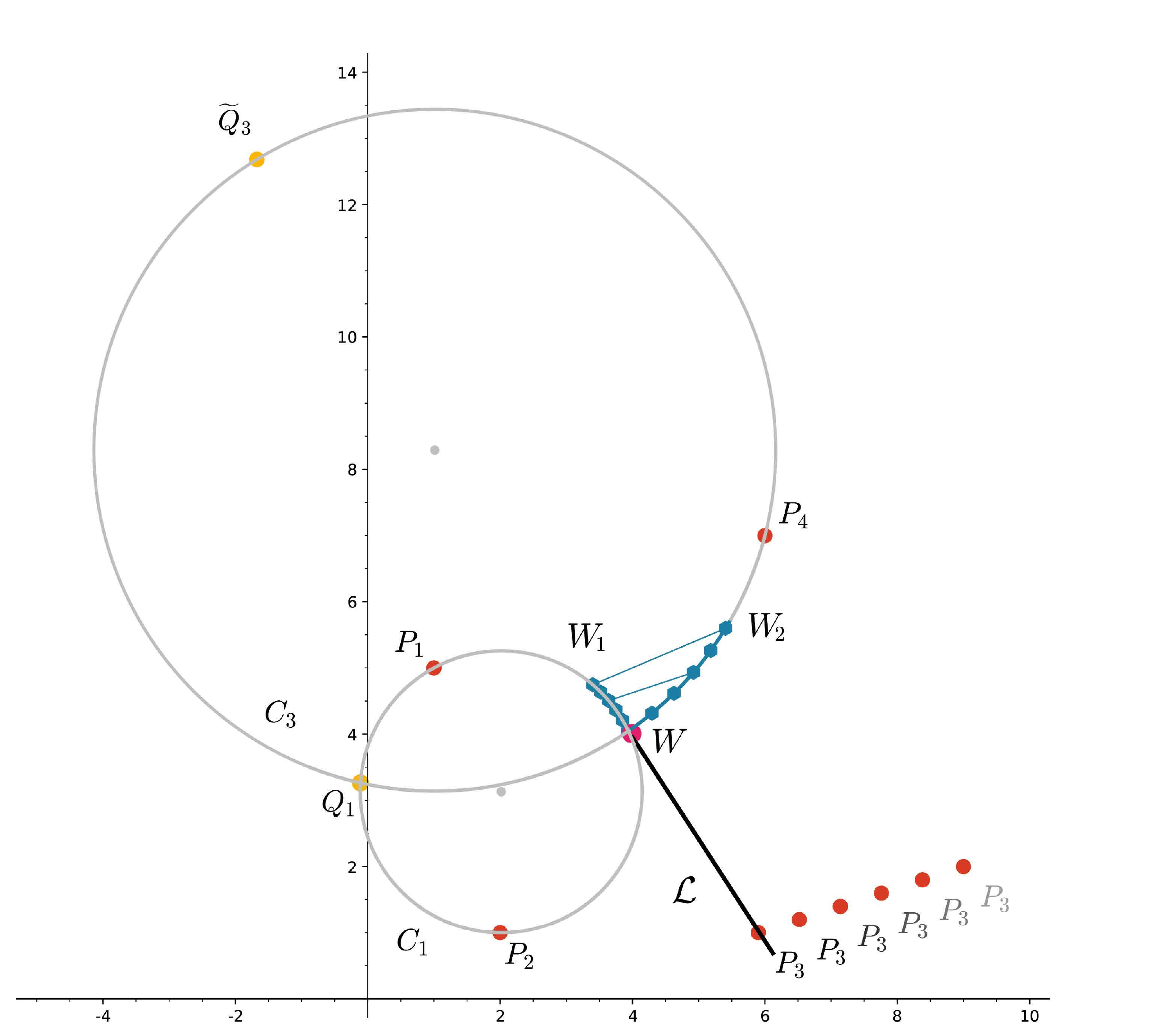}
\caption[]{Example \ref{ex2}. Dynamics of the facilities $ W_1 $ and $ W_2 $ under variation of the terminal $ P_3 $}\label{fig:moving_p3}
\end{figure}

%%%%%%%%%%%%%%%%%%%%%%%%%%%%%%%%%%%%%%%%%%%%%%%%%%%%%%
%  EXAMPLE 2
%%%%%%%%%%%%%%%%%%%%%%%%%%%%%%%%%%%%%%%%%%%%%%%%%%%%%%

At the same time, the facility $ W_2 $ drifts along the circle $ C_3 $  passing through the points
$ Q_1, P_4 $ and $ \widetilde Q_3 $ (Fig. \ref{fig:moving_p3}). Here $ Q_1 $ is defined as in solution to Example \ref{ft}, while  $ \widetilde Q_3 $ is constructed in the manner analogous to $ Q_1 $, i.e.
the triangle $ P_4 Q_1 \widetilde Q_3 $ should be similar to the weight triangle $ \{m_3,m_4,m\} $. Its coordinates can be obtained from Theorem \ref{Th3term3}:
$$
\widetilde Q_3=\left(\frac{235-12\sqrt{15} -36\sqrt{55}-5\sqrt{33}}{64},\
\frac{1020-9\sqrt{15} +149\sqrt{55}+60\sqrt{33}}{192} \right)
$$
$$
\approx (-1.674715, 12.681409) \, .
$$
The circle $ C_3 $ is given as
$$ (x-X_3)^2+ (y-Y_3)^2 = r_3^2  $$
where
$$
\begin{array}{lll}
X_3&=\frac{235}{64}-\frac{207}{880}\sqrt{55}-\frac{23}{704}\sqrt{33}-\frac{3}{16}\sqrt{15} & \approx 1.013521, \\
Y_3&= \frac{85}{16} +\frac{3427}{10560}\sqrt{55}+\frac{23}{176}\sqrt{33}-\frac{3}{64}\sqrt{15} & \approx 8.288416,\\
r_3&=\sqrt{\frac{32}{\sqrt{15}}+\frac{1808}{99}} \approx 5.150241  \, .
\end{array}
$$
The trajectory of $ P_3 $ does not influence those of $ W_1 $ and $ W_2 $, i.e.  both facilities do not leave the corresponding  arcs
for any drift of $ P_3 $ until the latter swashes the line $ \mathcal L= \widetilde  Q_3 W $. At this moment, $ W_1 $ collides with $ W_2 $ in the point
\begin{eqnarray*}
W &=& \Big(\frac{\scriptstyle{867494143740435}+114770004066285 \sqrt{33}+14973708000030 \sqrt{55}+19296850969306\sqrt{15}}{\scriptstyle{435004929875940}}, \\
& & \frac{\scriptstyle{581098602680450}+10154769229801\sqrt{15}+9689425113917 \sqrt{55}-18326585102850\sqrt{33}}{\scriptstyle{145001643291980}} \Big) \approx (3.936925, 4.048287)
\end{eqnarray*}
%$$
%\approx (3.936925, 4.048287)
%$$
which stands for the second point of intersection of the circles $ C_1 $ and $ C_3 $, and yields a solution to the unifacility Weber problem (\ref{Weber_uni}) for the configuration $ \left\{ \renewcommand{\arraystretch}{0.5}
\begin{array}{c|c|c|c} P_1 & P_2 & P_3 & P_4 \\ m_1 & m_2 & m_3 & m_4 \end{array} \right\} $
(due to Theorem \ref{Th4Inv}, location of $ W $ is invariant for any position of $  P_3 $ in $ \mathcal L $). The equation for $ \mathcal L $ is as follows:
\begin{eqnarray*}
y&=& \frac{\scriptstyle{2872083714}-841888053\sqrt{15}-546765794\sqrt{33}+411553980\sqrt{55}}{\scriptstyle{ 310250553}}x
+\frac{\scriptstyle{114568440}+1171692495\sqrt{15}+694024390\sqrt{33}-319467193\sqrt{55}}{\scriptstyle{620501106}}
\\
&\approx &  -1.538431\, x + 10.104975  \, .
\end{eqnarray*}
When $ P_3 $ crosses the line $ \mathcal L $, the solution to the bifacility Weber problem (\ref{F_Weber}) does not exist (while the unifacility counterpart (\ref{Weber_uni}) still possesses a solution).
\qed

\begin{theorem} \label{ThDyn1}
For any position of the terminal $ P_3 $, the facility $ W_1 $ lies in the arc of the circle $ C_1 $ passing through the points $ P_1, P_2 $ and $ Q_1=(q_{1x},q_{1y}) $ given by the formula (\ref{Q1}) where substitution $ m_3 \to m $ is made. At the same time, the facility $ W_2 $ lies in the arc of the circle $ C_3 $ passing through the points $ Q_1=(q_{1x},q_{1y}), P_4 $ and $ \widetilde Q_3 $. Here $ \widetilde  Q_3 $ is given by (\ref{Q1}) where substitution
$$
\begin{array}{c|c|c|c|c}
(x_1,y_1) & (x_2,y_2) & m_1 & m_2 & m \\
(x_4,y_4) & (q_{1x},q_{1y}) & m_4 & m & m_3
\end{array}
$$
is applied to.
\end{theorem}

The scenario for the facilities behaviour in Example \ref{ex2} looks similar to the equal weighted case (the Steiner problem) \cite{Uteshev_Steiner}, whereas the problem statement  of the next results is of a completely novel nature.

\begin{theorem} \label{ThDyn2} Let the circle $ C_1 $ and the point $ Q_1=(q_{1x},q_{1y}) $ be defined as in Theorem \ref{ThDyn1}.
For any value of the weight $ m_3 $, the optimal facility $ W_1 $ lies in the arc of the circle $ C_1 $. At the same time, the facility $ W_2 $ lies in the arc of the $ 4 $th degree algebraic curve passing through the points $ P_3, P_4 $ and $ Q_1 $. It is
 given by the equation
\begin{equation}
m^2 \left|\begin{array}{rrr} 1 & 1 & 1 \\ x & q_{1x} & x_3 \\ y & q_{1y} & y_3  \end{array}  \right|^2 \left[(x-x_4)^2+(y-y_4)^2\right] = m_4^2 \left|\begin{array}{rrr} 1 & 1 & 1 \\ x & x_3 & x_4 \\ y & y_3 & y_4  \end{array}  \right|^2  \left[(x-q_{1x})^2+(y-q_{1y})^2\right] \, .
\label{Varm_3}
\end{equation}
\end{theorem}

\begin{example}\label{ex211}
For the configuration
$$
\left\{\begin{array}{c|c|c|c|}
P_1=(1,5) & P_2=(2,1) & P_3=(7,2) & P_4=(6,7) \\
m_1=3 & m_2=2 & m_3 & m_4=7/2
\end{array} %\mbox{ and }
\ m=4
\right\} \, ,
$$
find the loci of the facilities $ W_1 $ and $ W_2 $
under variation of the weight $ m_3 $ within the interval $ [1.1, 6.5] $.
\end{example}

\begin{figure}[H]\center
\graphicspath{{Illustrations/}}
\includegraphics[scale=0.4]{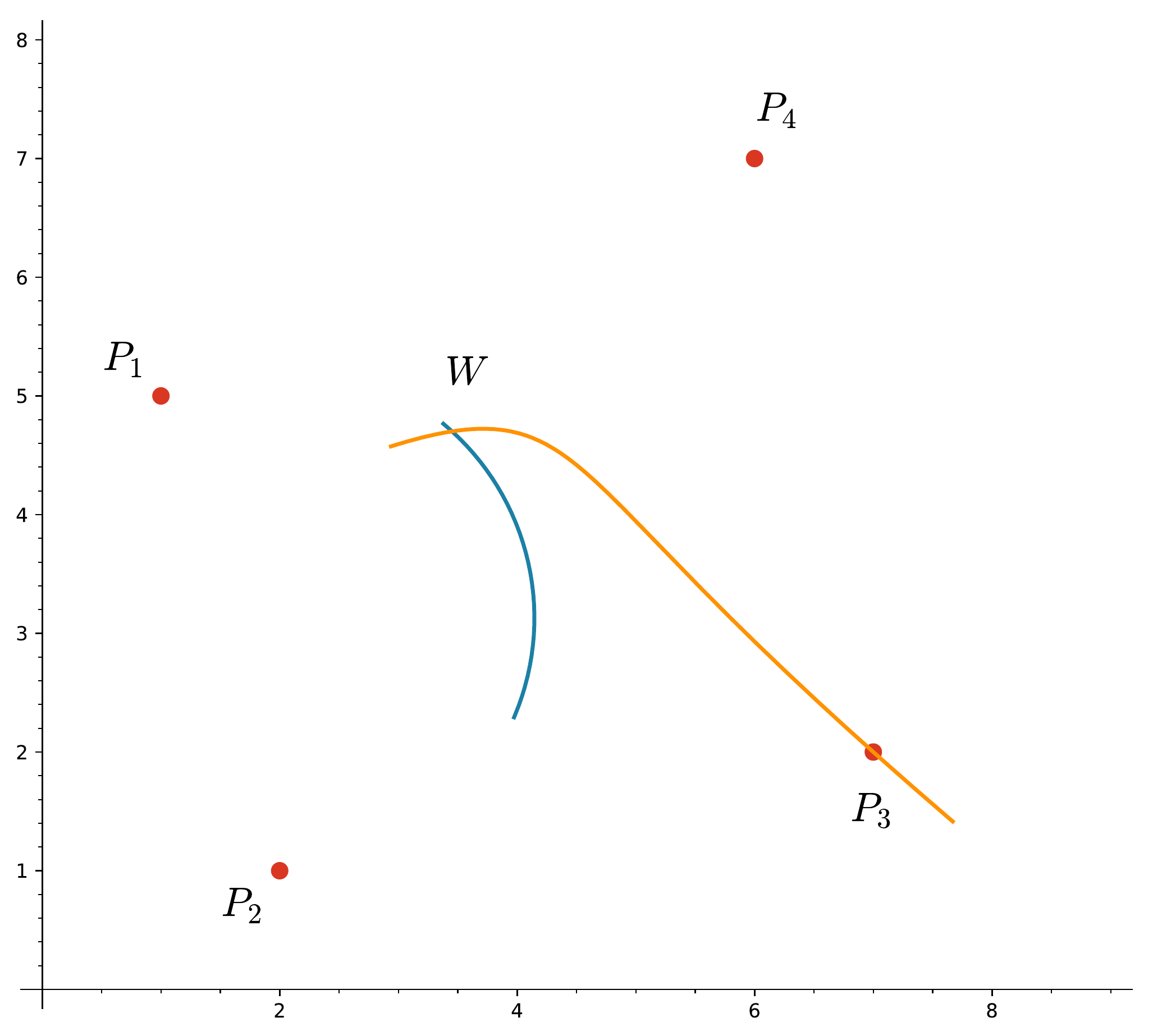}
\caption[]{Example \ref{ex211}. Dynamics of the facilities $ W_1 $ and $ W_2 $ under variation of the weight $ m_3 $.}\label{fig:moving_m_31}
\end{figure}

\textbf{Solution.} The facility $ W_1 $ moves along the arc of the circle $ C_1 $ given by (\ref{C1_ex3}).
The trajectory of $ W_2 $ is now a branch of the curve (\ref{Varm_3}) (Fig. \ref{fig:moving_m_31}) which we
 present here by its terms of the highest and the lowest degree
$$
(x^2+y^2)[7609x-(12924+597\sqrt{15})y][12243\,x+(13556+977\sqrt{15})y]
$$
$$
+\dots-143901885100-9445715749\sqrt{15}=0\, .
$$
It crosses that of $ W_1 $ when $ m_3 $ coincides with the zero of the equation $ \delta(m_3)=0 $. The latter
can be reduced to an algebraic one of the $ 8 $th degree. It happens to be even one in $ m_3 $ and, in principle, can be solved by radicals. We restrict ourselves here with a numerical approximation of this zero, namely  $ m_3=m_{3,1} \approx 1.394215 $.  The intersection point
$ W \approx (3.451796, 4.701666) $ yields a solution to the unifacility Weber problem for the configuration $ \left\{ \renewcommand{\arraystretch}{0.5} \begin{array}{c|c|c|c} P_1 & P_2 & P_3 & P_4 \\ m_1 & m_2 & m_{3,1} & m_4 \end{array} \right\} $ (v. solution of Example \ref{ex4t1f}).
In principle, the coordinates of $ W $ can also be expressed by radicals since the algebraic equations for their determination are of the $ 4 $th degree. When $ m_3 $ diminishes further from $ m_{3,1} $, the locus of $ W $ follows the curve found in the solution of Example \ref{ex4t1f}  and displayed in Fig. \ref{weber_531}. \qed

If the configuration of the previous example is slightly modified, the scenario for the network degeneracy varies.

\begin{example}\label{ex-2} For the configuration
$$
\left\{\begin{array}{c|c|c|c|}
P_1=(1,5) & P_2=(2,1) & P_3=(7,2) & P_4=(6,7) \\
m_1=3 & m_2=2 & m_3 & m_4=4
\end{array} %\mbox{ and }
 \ m=4  \right\} \, .
$$
find the loci of the facilities $ W_1 $ and $ W_2 $ under variation of the weight $ m_3 $ within the interval $ [0,7] $.
\end{example}

\textbf{Solution.} The trajectory of $ W_1 $ remains the same as in two previous examples. As for the facility $ W_2 $, this time the curve (\ref{Varm_3}) splits into the two algebraic curves: the cubic
$$
(x^2+y^2)(44809\,x+3183\,y\sqrt{15}+45696\,y)
$$
$$
+(8589\sqrt{15}-826932)x^2-(33588\sqrt{15}+850046)xy-(34053\sqrt{15}+793288)y^2
$$
$$
+(5231493-142194 \sqrt{15})x+
(5081366+177165\sqrt{15})y-12657634+489213\sqrt{15}=0
$$
and the line
\begin{equation}
3843\,x-(5184+283\sqrt{15})y-13230+1981\sqrt{15}=0 \, .
\label{ExLine}
\end{equation}
With $ m_3 $ decreasing from $ m_3=3 $ to $ 0 $, the facility $ W_1 $ moves towards $ P_1 $ while $ W_2 $ moves towards $ P_4 $ along the cubic.  These drifts tend to the points
$$ W_{1,0} = \left(\frac{332836}{253885}+\frac{305663}{761655}\sqrt{15}, \frac{788632}{253885}+\frac{129498}{253885}\sqrt{15} \right) \approx (2.865254, 5.081732),     $$
$$
W_{2,0} = \left(\frac{13871603}{3249728}+\frac{977199}{16248640}\sqrt{15}, \frac{18257483}{3249728}+\frac{1950153}{16248640}\sqrt{15} \right) \approx (4.501465,6.082990)
$$
correspondingly. It turns out that these points and $ P_4 $ lie in the line (\ref{ExLine}), and $ W_{1,0} $ is the solution to the unifacility Weber problem for the configuration $ \left\{ \renewcommand{\arraystretch}{0.5} \begin{array}{c|c|c} P_1 & P_2 & P_4 \\ m_1 & m_2 & m \end{array} \right\} $.

\begin{figure}[H]\center
\graphicspath{{Illustrations/}}
\includegraphics[scale=0.5]{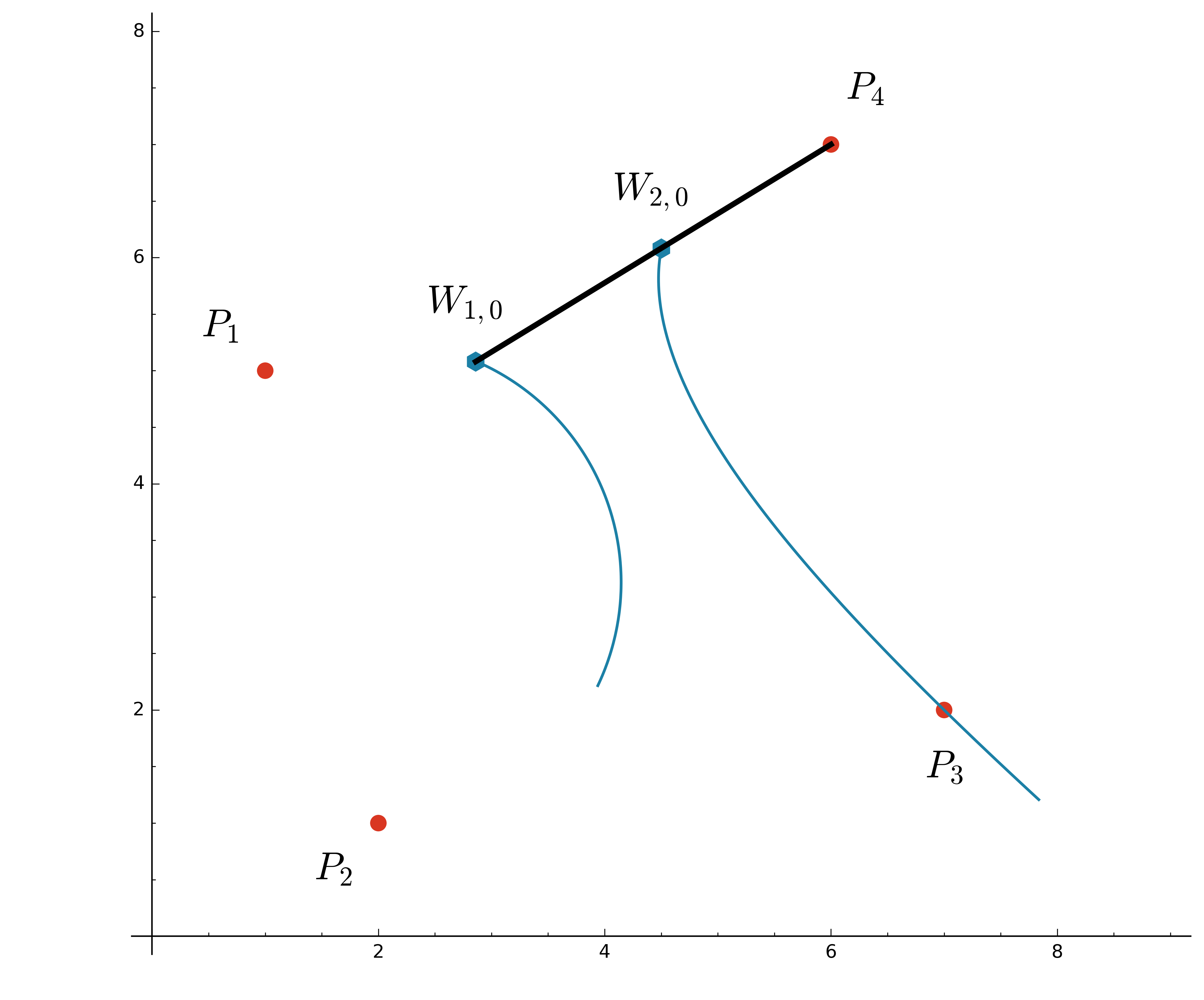}
\caption[]{Example \ref{ex-2}: Dynamics of the facilities $ W_1 $ and $ W_2 $ under variation of the weight $ m_3 $.}\label{fig:moving_m_3}
\end{figure}

With $ m_3 $ increasing from $ m_3=3 $, the facility $ W_1 $ moves to $ P_2 $ while $ W_2 $ moves to $ P_3 $.
Which terminal is reached faster? Due to (\ref{W1P1}), the answer depends on the relative position of the zeros of equations $ \delta_2(m_3)=0 $ and $ \delta_3(m_3)=0 $.
Via two successive squaring, both equations can be reduced to an algebraic form. The zero of $ \delta_2(m_3)=0 $ closest to $ m_3 = 3 $ is that of
$$
48\,841\,m_3^4-3\,283\,618\, m_3^2+19\,616\,041=0
$$
namely $ m_{3,2} \approx 7.784831 $. The zero of $ \delta_3(m_3)=0 $ closest to $ m_3 = 3 $ is that of
$$
48\,767\,485 m_3^8-6\,242\,238\,080 m_3^6+275\,224\,054\,560 m_3^4-4\,830\,235\,904\,000 m_3^2+28\,564\,663\,646\,464=0
$$
namely $ m_{3,3} \approx 6.607846 $. Therefore, one should first expect
the collision of $ W_2 $ with $ P_3 $ at $ m= m_{3,3} $ (Fig. \ref{fig:moving_m_3}). \qed

We finally treat the case of the variation of the parameter directly responsible for the inter-facilities connection.

%%%%%%%%%%%%%%%%%%%%%%%%%%%%%%%%%%%%%%%%%%%%%%%%%%%%%%
%  EXAMPLE 3
%%%%%%%%%%%%%%%%%%%%%%%%%%%%%%%%%%%%%%%%%%%%%%%%%%%%%%

\begin{example}\label{ex3} For the configuration
$$
\left\{\begin{array}{c|c|c|c|}
P_1=(1,5) & P_2=(2,1) & P_3=(7,2) & P_4=(6,7) \\
m_1=3 & m_2=2 & m_3=3 & m_4=4
\end{array} %\mbox{ and }
\ m \right\} \, ,
$$
find the loci of the facilities $ W_1 $ and $ W_2 $ under variation of the weight
$ m $ within the interval $ [2,4.8] $.
\end{example}

\begin{figure}[H]\center
\graphicspath{{Illustrations/}}
\includegraphics[scale=0.1]{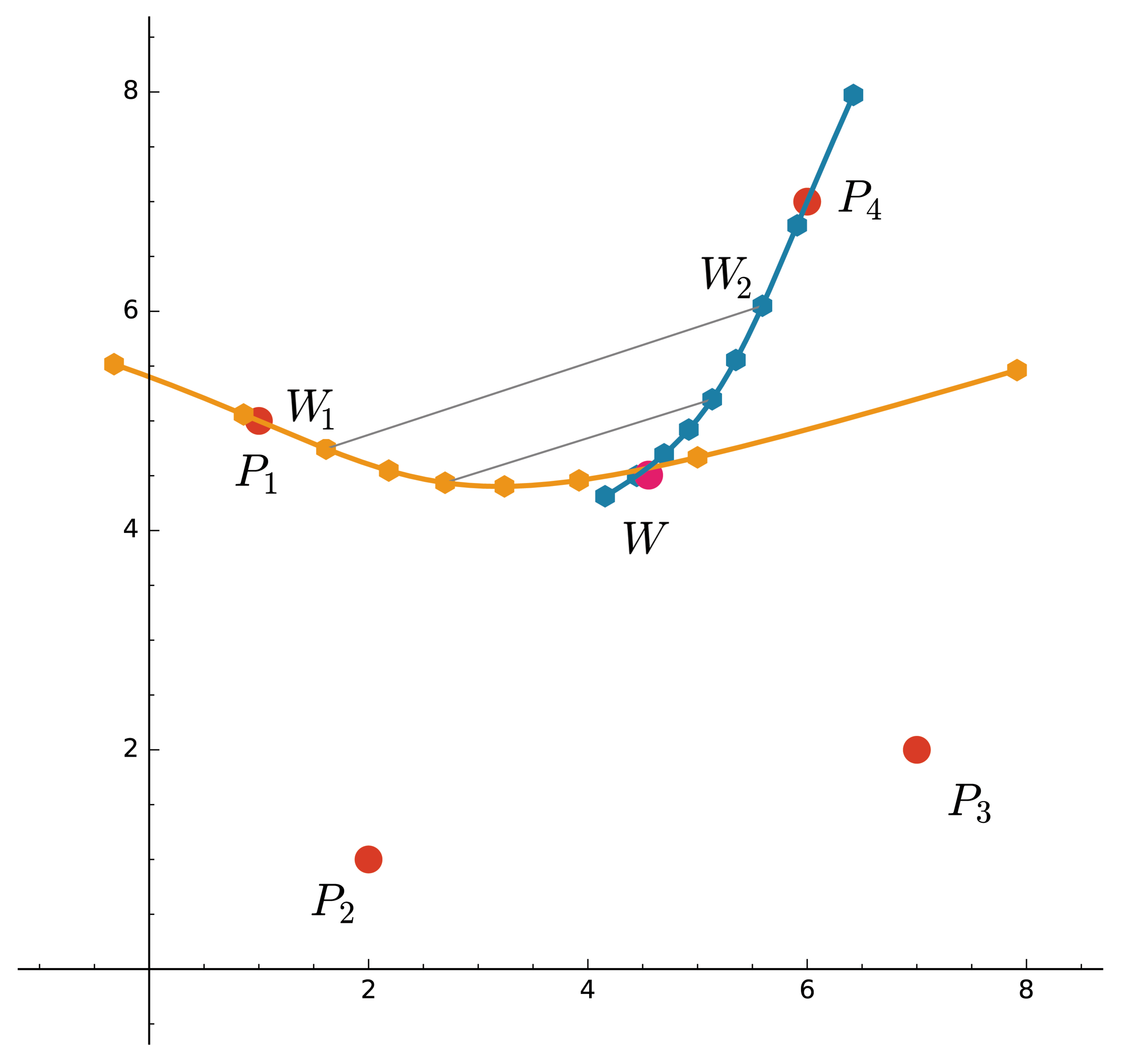}
\caption[]{Example \ref{ex3}: Dynamics of the facilities $ W_1 $ and $ W_2 $ under variation of the weight $ m $.}\label{fig:moving_m}
\end{figure}

\textbf{Solution.} When the weight $ m $ increases starting from $ m=4 $, the facilities $ W_1 $ and $ W_2 $ approach each other along the curves given in a parametric form as $ (x_{\ast}(m),y_{\ast}(m)) $ and $ (x_{\ast \ast}(m),y_{\ast \ast}(m)) $ correspondingly (Fig. \ref{fig:moving_m}). Using the resultant computation techniques, one can eliminate the parameter $ m $ and obtain the representation for both curves in an implicit form $ \Phi(x,y)=0 $ with a polynomial $ \Phi(x,y) $. We failed to deduce a general form for $ \Phi(x,y) $ for an arbitrary configuration (i.e., the counterpart of formula (\ref{Varm_3})). As for the configuration of the present example, the trajectory of $ W_1 $ follows the branch of the $12$th degree curve

\scriptsize
\begin{multline*}
(33x-64y)(57x-184y)(9x-4y)^2(x^2+y^2)^4
+2(9x-4y)(1197360\,y^4+38744xy^3-996250\,x^2y^2+177372\,x^3y+58743\,x^4)(x^2+y^2)^3\\
-({ 63907929}\,x^6-{ 341026596}\,x^5y+{ 357545379}\,x^4y^2-{ 144508432}\,x^3y^3+{ 467850026}\,x^2y^4+{ 183561728}\,xy^5-{ 216618832}\,y^6)(x^2+y^2)^2\\
-2({ 1409368256} y^7-{ 96009750}xy^6-{ 1257509320}x^2y^5-{ 1333223774}x^3y^4-{ 2414472761}x^4y^3+{ 56621289}x^5y^2+{ 60699935}x^6y+{ 201733101}x^7)\\
\times (x^2+y^2)+{ 2705648104}\,x^8-{ 6071141480}\,x^7y-{ 9161113572}x^6y^2-{ 31017248488}x^5y^3-{ 18764811475}x^4y^4-{ 13137751680}x^3y^5\\
+{ 18564788114} y^6 x^2+{ 13793961776}\, y^7 x+{ 23768272521}\, y^8+{ 4246051940}\,x^7+{ 35798248544}\,x^6y+{ 85312702272}\,x^5y^2+{ 9829041172}\,x^4y^3\\
-{ 69032921486}x^3y^4-{ 194526330228}x^2y^5-{ 166055023682}xy^6-{ 140446565264}\, y^7
-{ 50096988214} x^6-{ 130008320896}x^5y\\
+{ 68129615166}x^4y^2+{ 764574928104}x^3y^3+{ 1158538353585}x^2y^4+{ 1018036091148}xy^5+{ 613887715465}y^6\\
+{ 115264215460}\, x^5-{ 98538750356}\, x^4y-{ 2379471798756}\, x^3y^2-{ 4377187959972}\, x^2y^3-{ 3936257049738}\, xy^4-{ 2039180048286}\, y^5\\
- { 34219719043}x^4+{  2768991633800}x^3y+{  9624148447872}x^2y^2+{ 10219251620520}\,xy^3+{ 5137472010885}\, y^4\\
-{ 647188379554}\,x^3-{ 9899760752960}\, x^2y-{ 17194965296886}xy^2-{ 9574513269860}\, y^3+{ 2398452952559}\,x^2+{ 15421590991868}\, xy\\
+{ 12441150600039}\, y^2 -{ 3821653371050}\,x-{ 9498572135246}\,y+{ 2351359754194} = 0 \, .
\end{multline*}

\normalsize

Due to (\ref{W1W2}), the trajectories of $ W_1 $ and $ W_2 $ meet when $ m $ coincides with  a zero  of the equation $ \delta(m)=0 $. The latter can be reduced  to an algebraic one
$$
{\scriptstyle 24505}\, m^{20}-{\scriptstyle 3675750}\, m^{18}+{\scriptstyle 214114901}\,m^{16}
-{\scriptstyle 6100395704}\, m^{14}
+{\scriptstyle 88231771774}\, m^{12}-{\scriptstyle 596555669836}\,m^{10}
$$
$$
+{\scriptstyle 1454634503494}\,m^8-{\scriptstyle 2224914338408}\, m^6
+{\scriptstyle 13361952747497}\, m^4-{\scriptstyle 36933031029102}\, m^2 +{\scriptstyle 25596924755077}=0
$$
with a (closest to $ m=4 $) zero  $ m_{0,1} \approx 4.326092 $. The collision point $ W $ has its coordinates $ (x_{\ast},y_{\ast})  $ satisfying the  $ 10 $th degree
algebraic equations. Thus, for instance, $ x_{\ast} $ is a zero of the (irreducible over $ \mathbb Z $) equation
$$
{\scriptstyle 26172883257245641}\,x^{10}-{\scriptstyle 923131285793079898}\,x^9+{\scriptstyle 13307558344313669247}\,x^8-{\scriptstyle 96515684969701735656}\,x^7
$$
$$
+{\scriptstyle 299719922700274443198}\,x^6+{\scriptstyle 615788986911179454876}\,x^5-{\scriptstyle 10080764503138660399742}\,x^4+{\scriptstyle 47075811782361663673544}\,x^3
$$
$$
-{\scriptstyle 125596018219466046236391}\,x^2+{\scriptstyle 194932824845656435067806}\,x-{\scriptstyle 136815745438565609528929}=0 \, ,
$$
and $ x_{\ast} \approx 4.537574 $. The point $ W \approx (4.537574,4.565962 ) $ yields the solution to the unifacility Weber problem (\ref{Weber_uni}) for the configuration
$ \left\{ \renewcommand{\arraystretch}{0.5} \begin{array}{c|c|c|c} P_1 & P_2 & P_3 & P_4 \\ m_1 & m_2 & m_{3} & m_4 \end{array} \right\} $. This scenario demonstrates a paradoxical phenomenon: the weight  $ m $ increase forces the facilities to a collision, i.e. to
a network configuration where its influence disappears completely.

When $ m $ decreases from $ m=4 $, the facility $ W_1 $ moves towards $ P_1 $ while $ W_2 $ moves towards $ P_4 $. The first drift is faster than the second one: $ W_1 $ approaches $ P_1 $  when  $ m $ coincides with  a zero  of the equation $ \delta_1(m)=0 $. The latter can be reduced to an algebraic one
$$
 {\scriptstyle 377145}\,m^{12}-{\scriptstyle 15186678}\,m^{10}+  {\scriptstyle 245711056}\, m^8-{\scriptstyle 1983425640}\,m^6
 +{\scriptstyle 8079368573}\, m^4-{\scriptstyle 14857953930}\, m^2    + {\scriptstyle 8631109474}=0
$$
with a zero $ m_{0,2} \approx 3.145546 $.

%FIG 3

Let us finally watch the dynamics of the cost (\ref{cost}) of the optimal network when $ m $ increases.

\begin{figure}[H]\center
%\subfigure[Restoring $ P_1 $ and$ P_2 $ and solving the three terminal problem]
{
\graphicspath{{Illustrations/}}
\includegraphics[scale=0.3]{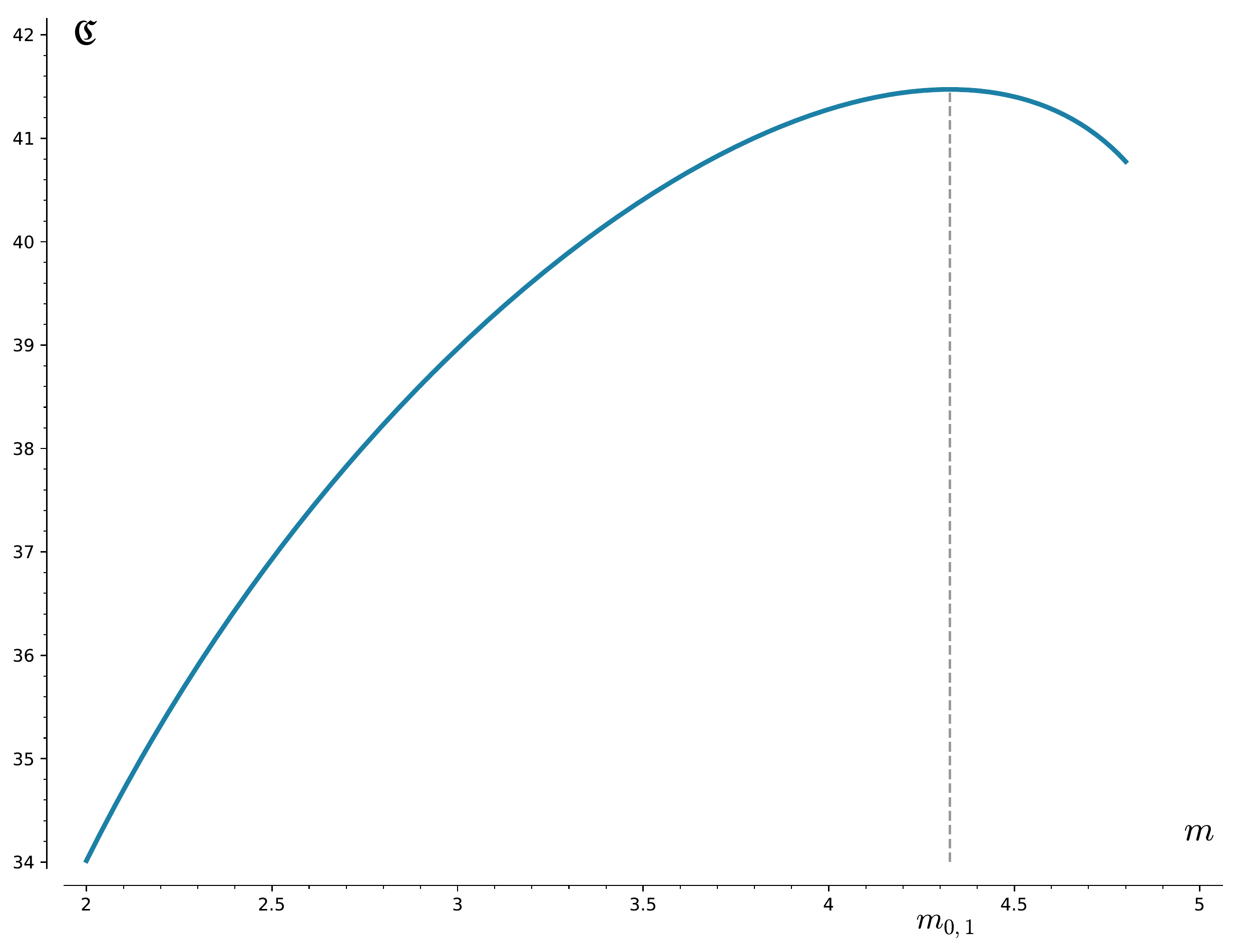}
   % \rule{4cm}{3cm}
    }
\caption{Example \ref{ex3}. Network cost as a function of the parameter $ m $. } \label{fig:cost}
\end{figure}

In Fig. \ref{fig:cost} one may notice that maximum value of $ \mathfrak C (m) $ is attained at the zero $ m_{0,1} \approx 4.326092 $ of $ \delta(m) $.
\qed

\begin{theorem}\label{Th_cost}
In the case of existence of the bifacility optimal network,  it is less costly than the unifacility one.
\end{theorem}

\textbf{Proof.} If the cost (\ref{cost}) is considered as the function of the configuration parameters then
the following identities are valid:
$$
\frac{\partial \mathfrak C^2}{\partial m_1 }\equiv \frac{m_1\delta_1}{m^2 \sqrt{\mathbf k_{12}}}\, ,\
\frac{\partial \mathfrak C^2}{\partial m_2 }\equiv\frac{m_2\delta_2}{m^2 \sqrt{\mathbf k_{12}}}\, ,\
\frac{\partial \mathfrak C^2}{\partial m_3 }\equiv \frac{m_3\delta_3}{m^2 \sqrt{\mathbf k_{34}}}\, ,\
\frac{\partial \mathfrak C^2}{\partial m_4 }\equiv \frac{m_4\delta_4}{m^2 \sqrt{\mathbf k_{34}}}
\quad \mbox{\rm and} \quad
\frac{\partial \mathfrak C^2}{\partial m }\equiv \frac{\delta}{2 m^3} \, .
$$
The last one results in
$$
\frac{\partial \mathfrak C}{\partial m} \equiv  \frac{\delta}{4m^3 \mathfrak C } \, .
$$
Therefore for any specialization of the weights $ \{m_j\}_{j=1}^4 $, the function  $ \mathfrak C(m) $ increases to its maximal value at the positive zero of $ \delta(m) $. \qed

Compared with the previous examples, in the solution of Example \ref{ex3} one cannot expect the coordinates of the point $ W $ to be expressed by radicals since the degrees of the resulted equations exceed $ 4 $.
Although this correlates somehow with the result by Bajaj \cite{Bajaj} that the unifacility Weber problem for the case of $ n>3 $ terminals is generically not solvable by radicals, further investigation of solvability by radicals in the general case should be carried out.

The empirical data obtained in the present section allows one to conclude that there are two possible ways of changing the bifacility topology of the optimal network to the unifacility one under variation of a configuration parameter. Collision of two facilities results in the appearance of a single facility with a valency equal to $ 4 $. Collision of a facility with a terminal or the scenario similar to that outlined in Example \ref{ex-2} keeps the valency of the remaining facility equal to $ 3 $.

%%%%%%%%%%%%%%%%%%%%%%%%%%%%%%%%%%%%%%%%%%%%%%%%%%%%%%
%%%%%%%%%%%%%%%%%%%%%%%%%%%%%%%%%%%%%%%%%%%%%%%%%%%%%%
%  5 TERMINALS
%%%%%%%%%%%%%%%%%%%%%%%%%%%%%%%%%%%%%%%%%%%%%%%%%%%%%%
%%%%%%%%%%%%%%%%%%%%%%%%%%%%%%%%%%%%%%%%%%%%%%%%%%%%%%

\section{Bifacility Case: Geometry Once Again}\label{S=4Alt}

In all the results of the present section we assume the fulfilment of conditions for existence of the solution to the bifacility Weber problem (\ref{F_Weber}) for the configuration
\begin{equation}
\left\{\begin{array}{c|c|c|c|}
P_1 & P_2 & P_3 & P_4 \\
m_1 & m_2 & m_3 & m_4
\end{array}\ m %\mbox{ and }
\right\} \ .
\label{direc_config}
\end{equation}

\begin{theorem} \label{Th4Alter} The facility $ W_2 $ lies in the line connecting the point $ \widetilde Q_3 $ defined in Theorem \ref{ThDyn1} with the terminal $ P_3 $. For the cost of the optimal network, one has
$$ \mathfrak C= m_3 | \widetilde Q_3 P_3 | \, . $$
\end{theorem}

This result gives rise to an alternative geometric construction for the facility points $ W_1 $ and $ W_2 $ in the optimal network. Namely, the facility $ W_2 $ is obtained as the point of intersection of the line $ \widetilde Q_3 P_3 $ with the circle $ C_3 $ defined also in Theorem \ref{ThDyn1}.  Therefore, Pick's geometric construction from Section \ref{SGeo} can be replaced by that consisting of the following steps:

\textbf{(a)} Construct the point $ Q_1 $ according to Pick's algorithm.

\textbf{(b)} Draw the circle $ C_1 $ through the points $ P_1, P_2 $ and $ Q_1 $.

\textbf{(c)} Construct the point $ \widetilde Q_3 $ defined in Theorem \ref{ThDyn1}.

\textbf{(d)} Draw the circle $ C_3 $ through the points $ P_4, Q_1 $ and $ \widetilde Q_3 $.

\textbf{(e)} Draw the line through $ \widetilde Q_3 $ and $ P_3 $; the intersection point with $ C_3 $ is $ W_2 $.

\textbf{(f)} Draw the line through $ Q_1 $ and $ W_2 $; the intersection point with $ C_1 $ is $ W_1 $.

The new algorithm ``costs'' $ 2 $ lines and $ 2 $ circles, and, compared with Pick's algorithm, one gets the one extra line construction excess. However the new algorithm is more suitable for dealing with a wandering terminal problem. Indeed, if the terminal $ P_3 $ is supposed to be moving somehow, the current optimal position of $ W_2 $ is obtained as the intersection point of the line passing through the fixed point $ \widetilde Q_3 $ with the invariant circle $ C_3 $.

Evidently, the suggested algorithm can be modified by attaching the construction to any terminal of the configuration other than $ P_3 $.
Let us construct, in a manner similar to $ \widetilde Q_3 $,
extra points $ \widetilde Q_1, \widetilde Q_2 $ and $ \widetilde Q_4 $. For the convenience of references, we place below the instructions for coordinates evaluation for all the points involved in construction. The starting point is
$$
Q_1=\Bigg(\frac{1}{2}(x_1+x_2)+\frac{(m_1^2-m_2^2)(x_1-x_2)-\sqrt{\mathbf k}(y_1-y_2)}{2m^2},  \frac{1}{2}(y_1+y_2)+\frac{(m_1^2-m_2^2)(y_1-y_2)+\sqrt{\mathbf k}(x_1-x_2)}{2m^2} \Bigg)
$$
with
$
\mathbf k=(m_1+m_2+m)(-m_1+m_2+m)(m_1-m_2+m)(m_1+m_2-m)
$.
The coordinates of the other points are obtained from these via replacement $ (P_1,P_2,m_1,m_2,m) $ by
$$
\begin{array}{c|c|c|c|c}
   (P_3,P_4,m_3,m_4,m) &  (P_2,Q_2,m_2,m,m_1) &  (Q_2,P_1,m,m_1,m_2) & (P_4,Q_1,m_4,m,m_3)  & (Q_1,P_3,m,m_3,m_4) \\
   \hline
  \mbox{\rm for} \ Q_2 &   \mbox{\rm for}\ \widetilde Q_1 & \mbox{\rm for}\ \widetilde Q_2 & \mbox{\rm for}\ \widetilde Q_3 & \mbox{\rm for}\ \widetilde Q_4
\end{array}
$$
Therefore, the triangle $ P_2Q_2\widetilde Q_1 $ is similar to the weight triangle $\{m_1,m_2,m \} $, etc.

\begin{figure}[H]\center
\graphicspath{{Illustrations/}}
\includegraphics[scale=0.8]{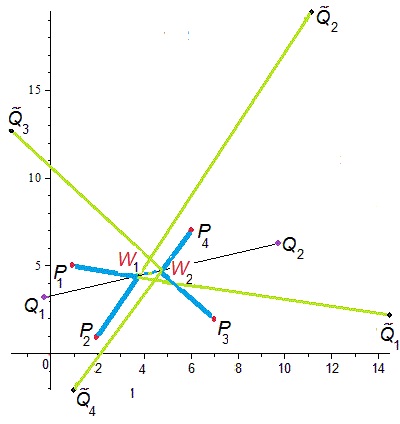}
\caption[]{Example \ref{ExDual}. Locus of the points $\widetilde Q_1,\widetilde Q_2,\widetilde Q_3,\widetilde Q_4 $. }\label{fig:dual_conf}
\end{figure}

\begin{example} \label{ExDual} For the configuration of Example \ref{ft} one gets (Fig. \ref{fig:dual_conf})
$$
\begin{array}{c|c|c|c|c}
&\widetilde Q_1 & \widetilde Q_2 & \widetilde Q_3 & \widetilde Q_4 \\
\hline
\approx & (14.46457,2.16301) & (11.16956, 19.49165) & (-1.67471,12.68140) & (1.01331,-2.03540)
\end{array}
$$
\end{example}

Thus the result of Theorem \ref{Th4Alter} can be extended to the following one

\begin{theorem} The facility $ W_2 $ lies in the line connecting the point $ \widetilde Q_4 $ with the terminal $ P_4 $.
The facility $ W_1 $ lies in the lines connecting the point $ \widetilde Q_1 $ with the terminal $ P_1 $ and $ \widetilde Q_2 $ with $ P_2 $.
 One has
$$ \left\{ \mathfrak C= m_j | \widetilde Q_j P_j | \right\}_{j=1}^4  \, . $$
\end{theorem}

To complete this section, we present an experimentally obtained mysterious conclusion which looks like a counterpart to Theorem \ref{Thdual}.

\textbf{Conjecture 1.} The configuration
\begin{equation}
\left\{\begin{array}{c|c|c|c|}
\widetilde Q_1 & \widetilde Q_2 & \widetilde Q_3 & \widetilde Q_4 \\
m_1 & m_2 & m_3 & m_4
\end{array}\ m %\mbox{ and }
\right\}
\label{dual_config}
\end{equation}
will be named the \textbf{dual configuration} for the configuration (\ref{direc_config}).
The values $ \delta $ and $ \widetilde \delta $ computed for these configurations via (\ref{del}) are related by the equality
$$ \widetilde \delta =- 3 \delta \, . $$
Therefore, due to Theorem \ref{teo1},  solution to the Weber problem (\ref{F_Weber}) for the configuration (\ref{dual_config}) does not exist. However,
formulae (\ref{W1x})--(\ref{W2y})  for the coordinates of the facilities formally applied to the configurations (\ref{direc_config})  and (\ref{dual_config}) yield the same facilities $ W_1 $ and $ W_2 $,
while the cost formula (\ref{cost}) results in the following equality
$$ \widetilde{\mathfrak C}= 3 \mathfrak C \, . $$

Of especial interest is the case of configurations satisfying the condition $ \delta=0 $. For this case, both the generating configuration and its dual configuration possess the same solution point to the unifacility Weber problem (\ref{Weber_uni}).

\begin{example} For the configuration
$$
\left\{\begin{array}{c|c|c|c|}
P_1=(1,5) & P_2=(2,1) & P_3=(7,2) & P_4=(6,7) \\
m_1=3 & m_2=2 & m_3=3 & m_4=4
\end{array} %\mbox{ and }
 \ m \right\}
$$
the bifurcation value of the weight $ m $ is evaluated in Example \ref{ex3} : $ m=m_{0,1}\approx 4.326092 $. For this value, the bifacility Weber problem becomes unsolvable, while the unifacility (i.e. the generalized Fermat-Torricelli) one (\ref{Weber_uni})  possesses the solution
$ W_{\ast} \approx (4.537574, 4.565962) $ with the cost of the network $ \mathfrak C \approx  41.473087 $. The dual network
$$
 \left\{\begin{array}{c|c|c|c|}
\widetilde Q_1 & \widetilde Q_2 & \widetilde Q_3 & \widetilde Q_4 \\
m_1 & m_2 & m_3 & m_4
\end{array}\ m_{0,1}
\right\}
$$
contains the following terminals
$$
\begin{array}{c|c|c|c|c}
&\widetilde Q_1 & \widetilde Q_2 & \widetilde Q_3 & \widetilde Q_4 \\
\hline
\approx & (14.72146,3.3164645) & (14.02292, 17.89537) & (-2.57199,11.97446) & 0.66019,-1.88749)
\end{array}
$$
while the points $ Q_1,Q_2 $ used in their construction are as follows
$$ Q_1 \approx (0.238805, 3.252425), Q_2 \approx (9.407076, 6.053893) \, .  $$

\textbf{(i)} All the lines $ \{P_j \widetilde Q_j\}_{j=1}^4 $ and $ Q_1Q_2 $ have the common point, namely $ W_{\ast} $. One has:
$$ \mathfrak C = m_{0,1} |Q_1Q_2|=m_j |P_j \widetilde Q_j| \quad \mbox{\rm for} \ j\in \{1,2,3,4\} \, .
$$
\textbf{(ii)} Fermat-Torricelli point for the configuration
$
 \left\{\begin{array}{c|c|c|c}
\widetilde Q_1 & \widetilde Q_2 & \widetilde Q_3 & \widetilde Q_4 \\
m_1 & m_2 & m_3 & m_4
\end{array}
\right\}
$
coincides with $ W_{\ast} $.

From \textbf{(i)} and \textbf{(ii)} it follows that the cost of this (unifacility) network equals
$ \widetilde{\mathfrak C} = 3 \mathfrak C \approx 124.419261 $.
\end{example}

\section{Five Terminals}\label{S>4}

 How is it possible to extend an analytical approach developed in Section \ref{SAn} to the multifacility Weber problem?
 For this aim, we once again address the geometric solution for the four-terminal problem from Section \ref{SGeo} but provide it with the alternative interpretation on the base of its analytical background. It turns out that the four-terminal Weber problem (\ref{F_Weber}) can be reduced
 to the pair of the three-terminal Weber problems. We will utilize abbreviations $ \{4\mathbf t 2\mathbf f\} $
and $ \{3{\mathbf t} 1\mathbf f\} $ for the cor\-res\-ponding problems.

Assume that solution for the $ \{4\mathbf t 2\mathbf f\} $-Weber problem (\ref{F_Weber}) exists. Then the system of equations (\ref{grad_42})-(\ref{grad_424}) providing the coordinates of the facilities could be split into two subsystems. Comparing  equations (\ref{grad_42}) and (\ref{grad_422}) with (\ref{grad_32}) and (\ref{grad_322}) permits one to claim that the optimal facility $ W_1 $ coincides with its counterpart for the $ \{3{\mathbf t} 1\mathbf f\} $-Weber problem for the configuration
$
\left\{ \renewcommand{\arraystretch}{0.5}
\begin{array}{c|c|c}
P_1 & P_2 & W_2 \\
m_1 & m_2 & m
\end{array}
\right\}
$.
A similar statement is also valid for the facility $ W_2 $, i.e. it is the solution to the Weber problem for the configuration
$
\left\{ \renewcommand{\arraystretch}{0.5}
\begin{array}{c|c|c}
P_3 & P_4 & W_1 \\
m_3 & m_4 & m
\end{array}
\right\} \, .
$
From this point of view, it looks like the four-terminal Weber problem can be reduced to the pair of the three-terminal ones.
However, this reduction should be modified since the loci of the facilities  $ W_2 $ or $ W_1 $ remain still undetermined. The result of Theorem \ref{Th3term3} permits one to replace these facilities by those with known positions.

\begin{theorem} \label{ThW5} If the solution to the $ \{4{\mathbf t} 2\mathbf f\} $-Weber problem (\ref{F_Weber}) exists then the facility $ W_2 $ coincides with the solution to the $ \{3{\mathbf t} 1\mathbf f\} $-Weber problem for the configuration $
\left\{ \renewcommand{\arraystretch}{0.5}
\begin{array}{c|c|c}
P_3 & P_4 & Q_1 \\
m_3 & m_4 & m
\end{array}
\right\}
%\label{split_conf21}
$.
A similar statement is valid for the terminal $ W_1 $: it coincides with the solution to the $ \{3{\mathbf t} 1\mathbf f\} $-Weber problem for the configuration
$
\left\{ \renewcommand{\arraystretch}{0.5}
\begin{array}{c|c|c}
P_1 & P_2 & Q_2 \\
m_1 & m_2 & m
\end{array}
\right\}
%\label{split_conf22}
$
Coordinates of points $ Q_1 $ and $ Q_2 $ are given in Section \ref{S=4Alt}.
\end{theorem}

This theorem claims that the four-terminal Weber problem can be solved by its reduction to the three-terminal counterpart via a formal replacement of a pair of the \textit{real} terminals, say $ P_3 $ and $ P_4 $, by a single \textit{phantom} terminal $ Q_2 $. This reduction algorithm is similar to that used for construction of the Steiner minimal tree (firstly introduced by Gergonne as early as in 1810, and 150 years later rediscovered by Melzak \cite{Brazil_2014}). The approach can be evidently extended to the general case of $ n \ge 5 $ terminals
as ia clarified by the following example.

%%%%%%%%%%%%%%%%%%%%%%%%%%%%%%%%%%%%%%%%%%%%%%%%%%%%%%
%  EXAMPLE
%%%%%%%%%%%%%%%%%%%%%%%%%%%%%%%%%%%%%%%%%%%%%%%%%%%%%%

\begin{example}\label{Ex5t3f}
Find the coordinates of the facilities $ W_1, \, W_2, \, W_3 $ that minimize the cost
\begin{equation}
m_1|P_1W_1|+m_2|P_2W_1|+m_3|P_3W_2| +  m_4|P_4W_2|+m_5|P_5W_3|
    +  \widetilde{m}_{1,3} |W_1W_3| + \widetilde{m}_{2,3} |W_2W_3|
\label{cost5-3}
\end{equation}
for the following configuration:
$$
\left\{\begin{array}{c|c|c|c|c|c}
P_1=(1,6) & P_2=(5,1) & P_3=(11,1) & P_4=(15,3) & P_5=(7,11) & \widetilde{m}_{1,3}=10  \\
m_1=10 & m_2=9 & m_3=8 & m_4=7 & m_5=13 & \widetilde{m}_{2,3}=12
\end{array}
\right\} \, .
$$
\end{example}

\textbf{Solution.} \textbf{(I)} To reduce the problem to the $ \{4{\mathbf t} 2\mathbf f\} $-case, replace a pair of the terminals $ P_1 $ and $ P_2 $ by the point $ Q_1 $ defined by the formula (\ref{Q1}) where the substitution $ m_3 \to \widetilde{m}_{1,3} $ is made.
$$
Q_1=\left( -\frac{9}{40}\sqrt{319} + \frac{131}{50} , -\frac{9}{50}\sqrt{319} + \frac{159}{40} \right) \approx ( -1.398628 , 0.760097) \, .
$$

\textbf{(II)} Solve the $ \{4{\mathbf t} 2\mathbf f\} $-problem for the configuration
$\left\{ \renewcommand{\arraystretch}{0.5} \begin{array}{c|c|c|c|}
P_5 & Q_1 & P_3 & P_4 \\
m_5    & \widetilde{m}_{1,3} & m_3 & m_4
\end{array}  \widetilde{m}_{2,3}
\right\}
$
via formulae \eqref{W1x}--\eqref{W2y}  and obtain the coordinates for the facilities
$$
W_2 \approx ( 10.441211 , 3.084533 )
 \ \mbox{ and } \ W_3\approx (7.191843 , 5.899268 ) \, .
$$

\begin{figure}[H]\center
%\subfigure[Restoring $ P_1 $ and$ P_2 $ and solving the three terminal problem]
{
\graphicspath{{Illustrations/}}
\includegraphics[scale=0.5]{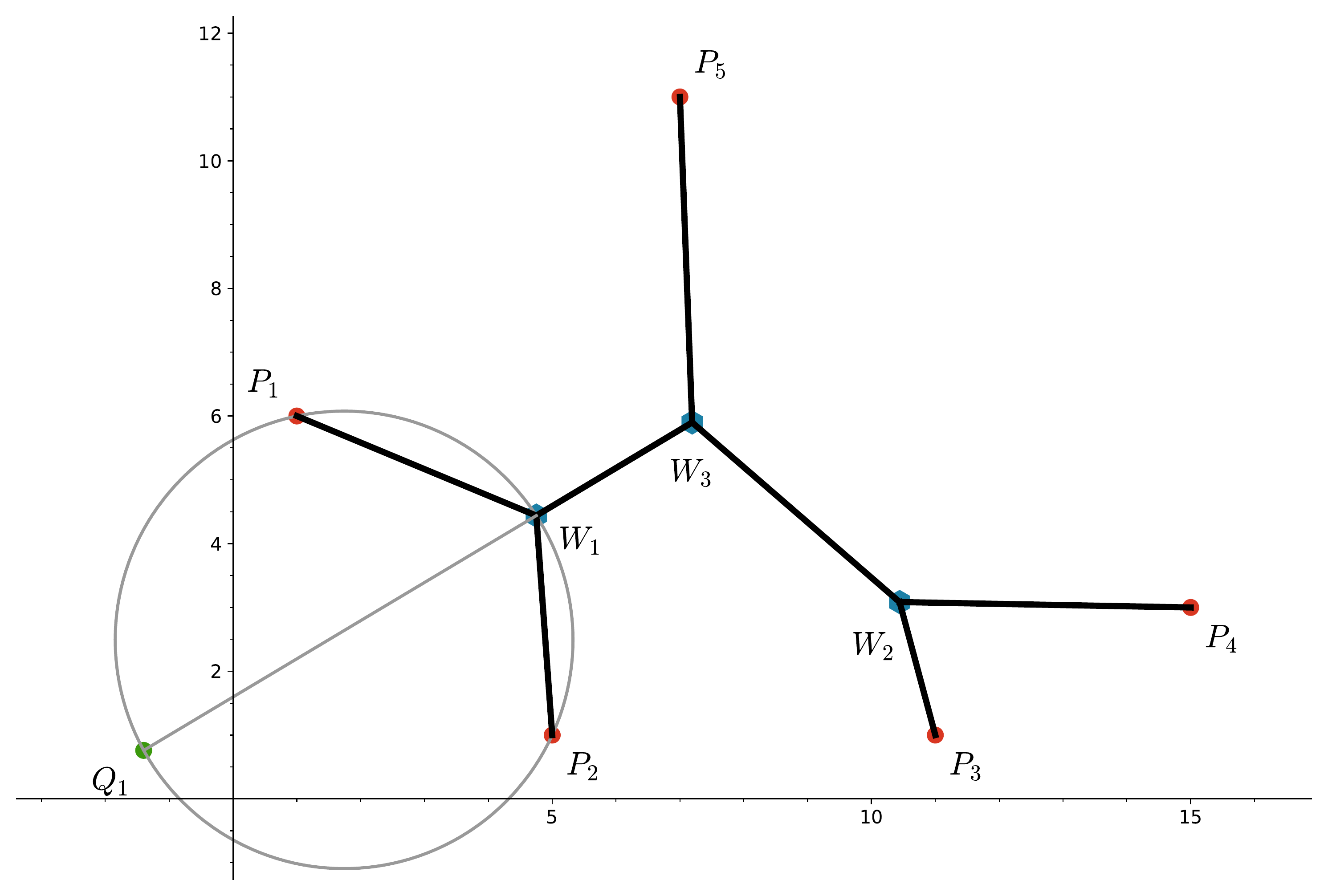}
   % \rule{4cm}{3cm}
    }
\caption{Example \ref{Ex5t3f}. Weber network construction for five terminals} \label{fig:5-3}
\end{figure}

\textbf{(III)} Return $ P_1 $ and $ P_2 $  instead of $ Q_1 $ and solve the $ \{3{\mathbf t} 1\mathbf f\} $-Weber problem for the configuration
$
\left\{ \renewcommand{\arraystretch}{0.5}
\begin{array}{c|c|c}
P_1 & P_2  & W_3 \\
m_1 & m_2 & \widetilde{m}_{1,3}
\end{array}
\right\}
$
by the formulae of Theorem \ref{Th3term1}: $ W_1\approx (4.750727 , 4.438893) $ (Fig. \ref{fig:5-3}). We emphasize, that the coordinates of the facilities can be expressed by radicals similar to the following expression for the cost of the network
$$
\mathfrak C =\frac{\sqrt{10}}{80} \big( 4158\sqrt{87087}+773402\sqrt{231}+271890\sqrt{319}+247470\sqrt{143}+326403\sqrt{609}
$$
$$
+104181\sqrt{273}-4455\sqrt{377}+15216515 \big)^{1/2} \approx 267.229644  \ .
$$
\qed

%\begin{figure}[H]
%\subfigure[Construction of an adjuvant point $Q$]{
%\graphicspath{{Illustrations/}}
%\includegraphics[scale=0.56]{5-3_1}
%    %\rule{4cm}{3cm}
%    \label{fig:5-3_1}
%    }
%\subfigure[Reducing the problem to the four terminal case via replacing the pair $ P_1, P_2 $ by phantom terminal $ Q_1 $]{
%\graphicspath{{Illustrations/}}
%\includegraphics[scale=0.56]{5-3_2}
%   % \rule{4cm}{3cm}
%   \label{fig:5-3_2}
%}
%\end{figure}
%

The reduction procedure illuminated in the previous example, in the general case should be accompanied by the conditions similar to those from Theorem \ref{teo1}.

We conclude this section with formulation of two problems for further research. For the background  of the first one, we refer to the last example. When the weights $ \widetilde{m}_{1,3} $ and $ \widetilde{m}_{2,3} $ increase, the optimal facilities tend to the collision point $ W\approx (6.699788,4.813544) $ that coincides with the generalized Fermat-Torricelli point for the configuration
$\left\{ \renewcommand{\arraystretch}{0.5} \begin{array}{c|c|c|c|c}
P_1 & P_2 & P_3 & P_4 & P_5\\
m_1    & m_2 & m_3 & m_4 & m_5
\end{array}
\right\}
$. Therefore, there exists a lower bound for the sum $ \widetilde{m}_{1,3} + \widetilde{m}_{2,3} $ such that the cost of the corresponding optimal solution to the unifacility Weber problem is less than that of any $ 3 $-facilities  configuration of the type (\ref{cost5-3}).

Using the terminology of the mechanics of materials we pose the following

\textbf{Problem of ultimate tensile weight.} Let there exists a solution  to the unifacility problem (\ref{Weber_uni}) (not coinciding with any $ P_j $). Find the minimal possible value for the sum
$$ \sum_{k=1}^{\ell} \sum_{i=k+1}^{\ell-1} \widetilde m_{ik} $$
such that the cost of this  solution is less than the cost
$$
\sum_{j=1}^n \sum_{i=1}^{\ell} m_{ij} |W_iP_j| +
\sum_{k=1}^{\ell} \sum_{i=k+1}^{\ell-1} \widetilde m_{ik} |W_iW_k|
$$
of any $ \ell $-facility network for arbitrary $ \ell>1 $.

\begin{example} Let the quadrilateral $ P_1P_2P_3P_4 $ be convex. For the equal weighted configuration
$\left\{ \renewcommand{\arraystretch}{0.5} \begin{array}{c|c|c|c}
P_1 & P_2 & P_3 & P_4 \\
1    & 1 & 1 & 1
\end{array}
\right\} \, ,
$
the ultimate tensile weight equals
$$ 2 \max \left\{ \sin \psi/2 , \cos  \psi/2 \right\} $$
where $ \psi $ is the angle between the diagonal vectors $ \overrightarrow{P_1P_3} $ and $ \overrightarrow{P_2P_4} $. Evidently, this weight is within the interval $ [\sqrt{2},2] $.
\end{example}

The potential existence of the optimal networks containing facilities being the endpoints of more than $ 3 $ edges, significantly complicate solution to the Weber problem. This occasion essentially distinguishes the problem from that of Steiner minimal tree construction. However, due to the evident similarity of the two problems and with reference to the above obtained results, we announce the following

\textbf{Conjecture 2.} The $ \{n \ \mathbf{terminals}\ \ell\ \mathbf{facilities}\} $-Weber problem (\ref{F_Weber_m}) is solvable by radicals if $ \ell = n-2 $ and the valency of every facility in the network equals $ 3 $.

\section{Conclusions}

We provide an analytical solution to the bifacility Weber problem (\ref{F_Weber})  approving thereby the geometric
solution by G.Pick. We also formulate the conditions for the existence of the  network in a prescribed topology
and analyze the potential scenarios of its degeneracy under variation of parameters.

Several problems for further investigations are mentioned in Sections \ref{SPar}, \ref{S=4Alt} and \ref{S>4}. One extra problem
concerns the treatment of distance depending functions like $ F_L(P)= \sum_{j=1}^n m_j |PP_j|^L $ with different exponents $ L \in \mathbb Q \setminus 0 $.
 Specialization $ L=-1 $ corresponds to the Newton or the Coulomb potential. It turns out that the stationary point sets of all the functions $ \{F_L\}  $ can be treated
in the universal manner \cite{Uteshev_FTC}. We hope to discuss these issues in the foregoing papers.

%\section{Acknowledgement}

%This research was supported by the RFBR according to the project No~\textbf{17-29-04288}.

%%%%%%%%%%%%%%%%%%%%%%%%%%%%%%%%%%%%%%%%%%%%%%%%%%%%%%
%%%%%%%%%%%%%%%%%%%%%%%%%%%%%%%%%%%%%%%%%%%%%%%%%%%%%%
%  bibliography
%%%%%%%%%%%%%%%%%%%%%%%%%%%%%%%%%%%%%%%%%%%%%%%%%%%%%%
%%%%%%%%%%%%%%%%%%%%%%%%%%%%%%%%%%%%%%%%%%%%%%%%%%%%%%

\bibliographystyle{model1-num-names}
\nocite{*}

\section*{\refname}
\bibliography{mybib}

\end{document}